\documentclass{aa}  

\usepackage{lscape}
\usepackage{hyperref}
\bibpunct{(}{)}{;}{a}{}{,} 
\usepackage{graphicx}
\usepackage{txfonts}
\usepackage{hyperref}
\hypersetup{
    colorlinks=true,
    linkcolor=blue,
    filecolor=blue,      
    urlcolor=blue,
    citecolor=blue
    }
\newcommand{\Macc}{\mathrm{M_\odot \, yr^{-1}}}

\begin{document}

   \title{Modeling the secular evolution of embedded protoplanetary discs}

   \author{J. Mauxion
          \and
          G. Lesur
          \and 
          S. Maret
          }
   \institute{Univ. Grenoble Alpes, CNRS, IPAG, 38000 Grenoble, France\\
              \email{jonah.mauxion@univ-grenoble-alpes.fr}}

   \date{Received October 27, 2023; accepted March 25, 2024}

 
  \abstract
   {Protoplanetary discs are known to form around nascent stars from their parent molecular cloud as a result of angular momentum conservation. As they progressively evolve and dissipate, they also form planets. While a lot of modeling efforts have been dedicated to their formation, the question of their secular evolution, from the so-called class 0 embedded phase to the class II phase where discs are believed to be isolated, remains poorly understood.
   }
   {We aim to explore the evolution between the embedded stages and the class II stage. We focus on the magnetic field evolution and the long-term interaction between the disc and the envelope.}
   {We use the GPU-accelerated code \textsc{Idefix} to perform a 3D, barotropic, non-ideal magnetohydrodynamic (MHD) secular core collapse simulation that covers the system evolution from the collapse of the pre-stellar core until $100 \, \mathrm{kyr}$ after the first hydrostatic core formation and the disc settling while ensuring sufficient vertical and azimuthal resolutions (down to $10^{-2}$ au) to properly resolve the disc internal dynamics and non-axisymmetric perturbations.}
   {The disc evolution leads to a power-law gas surface density in Keplerian rotation that extends up to a few $10 \, \mathrm{au}$. The magnetic flux trapped in the disc during the initial collapse decreases from $100 \, \mathrm{mG}$ at disc formation down to $1 \, \mathrm{mG}$ by the end of the simulation. After the formation of the first hydrostatic core, the system evolves in three phases. A first phase with a small ($\sim 10 \, \mathrm{au}$), unstable, strongly accreting ($\sim10^{-5} \, \Macc$) disc that loses magnetic flux over the first $15 \, \mathrm{kyr}$, a second phase where the magnetic flux is advected with a smooth, expanding disc fed by the angular momentum of the infalling material, and a final phase with a gravitationally-regulated $\sim 60\,\mathrm{au}$ disc accreting at at few $10^{-7} \, \Macc$. The initial isotropic envelope eventually feeds large-scale vertically-extended accretion streamers, with accretion rates similar to that onto the protostar ($\sim 10^{-6} \, \Macc$). Some of the streamer material collides with the disc's outer edge and produces accretion shocks, but a significant fraction of the material lands on the disc surface without producing any noticeable discontinuity.
   }
   {While the initial disc size and magnetisation are set by magnetic braking, self-gravity eventually drives accretion, so that the disc ends up in a gravitationally-regulated state. This evolution from magnetic braking to self-gravity is due to the weak coupling between the gas and the magnetic field once the disc has settled. The weak magnetic field at the end of the class I phase ($B_z\sim 1 \, \mathrm{mG}$) is a result of the magnetic flux dilution in the disc as it expands from its initial relatively small size. This expansion should not be interpreted as a viscous expansion, as it is driven by newly accreted material from large-scale streamers with large specific angular momentum.}


   \maketitle
%

\section{Introduction}
\label{sec:intro}

Protoplanetary discs are ubiquitous in star-forming systems. Once they have formed, they are believed to be the main reservoir of mass that feeds the protostar and forms planets. In the early stages of their evolution, they are still embedded in a massive infalling envelope. As the system evolves, the envelope is progressively accreted onto the disc, which acts as a buffer between the envelope and the protostar. It is, therefore, crucial to understand the long-term evolution of such discs. They likely result from a complex interplay between mass input from the envelope and mass removal through accretion onto the protostar, outflowing material, and planet formation.

In essence, protoplanetary discs are the result of angular momentum conservation in a process that gathers mass from a $10^3 \, \mathrm{au}$ scale down to a few tens of $\mathrm{au}$. The initial configuration and later evolution of the disc are set by the amount of angular momentum stored in the gas by the time of its formation, and the mechanisms that are able to modify this amount.

On the one hand, it is now clear from core-collapse models of growing physical complexity \citep[see][for a review]{tsukamoto2022role} that we must both account for the initial magnetisation of the pre-stellar core and its complex chemistry to consistently reproduce the range of sizes and masses inferred in young discs from observational surveys \citep{maury2019characterizing,maret2020searching,tobin2020vla,sheehan2022vla}. Yet, robust conclusions about the relative importance of these ingredients and the influence of the initial conditions are still lacking.

On the other hand, results from numerical models emphasize the importance of magnetisation and self-gravity in the disc formation and early evolution. \citet{masson2016ambipolar} find that ambipolar diffusion is crucial during the collapse and for disc formation, as it decouples the magnetic field and the gas sufficiently to prevent the magnetic braking catastrophe, that suppresses the disc in ideal MHD simulations \citep{price2007effect}. \citet{hennebelle2016magnetically} support this result and show that the size of the newborn disc is set through a balance between magnetic braking (driven by toroidal magnetic field amplification) and ambipolar diffusion.

This magnetically-regulated phase is often followed by a gravitationally-regulated one. In their work, \citet{tomida2017grand,xu2021formation1,xu2021formation2} find a disc that becomes so massive and resistive that it is mainly controlled by angular momentum transport induced by self-gravity. In such a situation, they argue that the disc is stuck in a feedback loop where mass influx from the envelope promotes the generation of self-gravitating density waves that heat the gas, thus stabilizing the disc. Hence, while we have a good understanding of the relative importance of magnetisation and self-gravity during the early disc evolution, their influence on a secular scale remains to be explored.

As it falls onto the disc, the envelope can provide additional angular momentum and promote disc growth. The classic picture of an isotropic infall through a flattened, pressure-supported envelope (also known as pseudo-disc) is questioned by recent core collapse simulations of sufficiently massive molecular cloud accounting for turbulent, non-ideal magnetohydrodynamics \citep[MHD,][]{kuffmeier2017zoom, kuznetsova2019origins,lebreuilly2021protoplanetary}. In these simulations, the infall from the envelope is anisotropic and takes the form of filamentary or sheet-like structures. Such structures are reminiscent of the large-scale accretion streamers observed these recent years in embedded systems \citep{yen2019hl, pineda2020protostellar, murillo2022cold,valdivia2022prodige}

Hence, to understand the secular evolution of protoplanetary discs, one must understand the accretion mechanisms at stake in the disc. Magnetic braking, controlled by diffusion effects and the magnetic field intensity, is the best candidate. Yet, in cases where it becomes inefficient, self-gravity takes over. Thus, understanding the secular evolution of the magnetic field is key to understanding the regulation processes of the disc. The field may also play a role in the formation of anisotropies in the envelope.

Modeling the formation and evolution of protoplanetary discs is challenging because of the complex physics, the implied ranges of size and time, and the tri-dimensional nature of some key processes. For this reason, most collapse models either under-resolve the disc vertical structure or limit the computation time to the early class I stage. Yet, \citet{lubow1994magnetic} showed that the magnetic field radial diffusion efficiency depends on the disc thickness. It is also important to correctly capture phenomena such as magnetic field line twisting or spiral density wave generation. Thus, it is crucial to properly resolve the disc vertical extent while integrating for a significant time after the disc formation to study the disc and field evolution on a secular scale.

This paper is organized as follows: in Sect.~\ref{sec:method} we present our method and numerical setup, as well as the initial conditions of our model. Section~\ref{sec:overall} follows the overall evolution of the setup, starting from the isothermal collapse and browsing the secular evolution of the disc. In Sect.~\ref{sec:fidAccretion} we draw the accretion history of the disc as a complex interplay between magnetic braking, gravitational instability, and angular momentum influx from the envelope while Sect.~\ref{sec:streamer} probes the disc-envelope interaction that arises in the form of a large-scale accretion streamer. Finally, Sect.~\ref{sec:discussion} confronts our results with observational and numerical constraints. We conclude in Sect.~\ref{sec:conclusion}.

\section{Method}
\label{sec:method}

We aim at performing a long timescale core collapse simulation using the finite volume code \textsc{Idefix} \citep{lesur2023idefix}. This section presents the code and setup general properties.

\subsection{Governing equations and integration scheme}

The framework of our simulation lies in the context of non-relativistic, non-viscous, and locally isothermal non-ideal magnetohydrodynamics (MHD). The code solves for the classic mass, momentum, and Maxwell's equations:

\begin{align}
    \label{eq:mass}&\partial_t \rho + \mathbf{\nabla} \cdot (\rho \mathbf{u}) = 0,\\
    &\label{eq:momentum}\partial_t (\rho \mathbf{u}) + \mathbf{\nabla}\cdot (\rho \mathbf{u}  \otimes \mathbf{u}) = -\mathbf{\nabla}P-\rho\mathbf{\nabla}\phi_{g} + \frac{\mathbf{J}\times\mathbf{B}}{c},\\
    &\label{eq:induction}\partial_t\mathbf{B} = - \mathbf{\nabla}\times \mathbf{\mathcal{E}},\\
    &\label{eq:divB} \mathbf{\nabla}\cdot \mathbf{B} = 0,\\
    &\label{eq:current} \mathbf{J} = \frac{c}{4\pi}\mathbf{\nabla}\times \mathbf{B},
\end{align}

\noindent
where $\rho$, $P$, $\mathbf{u}$, $\mathbf{B}$ and $\mathbf{J}$ are respectively the density, the thermal pressure, the gas velocity, the magnetic field threading the medium and the electrical current. $c$ is the speed of light and $\phi_g$ is the gravitational potential. It is the sum of a point mass contribution from central mass $\phi_{pm}$ (see Appendix.~\ref{app:gstep}) and a self-gravitational contribution $\phi_{sg}$ which is connected to the density distribution via the Poisson equation:

\begin{equation} \label{eq:poisson}
    \Delta \phi_{sg} = 4\pi G \rho
\end{equation}

\noindent
with $G$ the gravitational constant. We assume the point mass to be fixed at the center and therefore neglect any reaction to the accretion of non-zero linear momentum material. The electromotive field $\mathbf{\mathcal{E}}$ is derived from the non-ideal Ohm's law in the case of ambipolar and Ohmic diffusions and reads:

\begin{equation} \label{eq:ohm}
    \mathbf{\mathcal{E}} = -\mathbf{u}\times\mathbf{B}+\frac{4\pi}{c}\eta_O\mathbf{J}-\frac{4\pi}{c}\eta_A\mathbf{J}\times\mathbf{b}\times\mathbf{b}
\end{equation}

\noindent
where $\eta_O$ and $\eta_A$ are the Ohmic and ambipolar diffusion coefficients, and $\mathbf{b}$ is a unitary vector aligned with the magnetic field.

\textsc{Idefix} solves the above equations using a conservative Godunov method \citep{toro2009hll} with a Constrained Transport (CT) scheme to evolve the magnetic field \citep{evans1988simulation}. The parabolic terms associated with non-ideal effects are computed separately using a super-timestepping Runge-Kutta-Legendre scheme \citep{meyer2014stabilized,vaidya2017scalable}. To prevent the accumulation of round-off errors on $\mathbf{\nabla}\cdot \mathbf{B}$ induced by the super-timestepping, we use a modified CT scheme in which the primitive variable evolved by the code is the vector potential $A$ on cell edges in place of the magnetic field $B$ on cell faces, as recommended by \citet{lesur2023idefix}. We implemented a biconjugate gradient stabilized (BICGSTAB) method with preconditioning that iteratively solves the Poisson equation (see Appendix~\ref{app:sgsolver} and Appendix~\ref{app:gstep} for the method and its application to our problem). Finally, a grid coarsening procedure \citep{zhang2019conservative} is applied in the azimuthal direction close to the axis to increase the integration timestep without loss of resolution in the equatorial region.

\subsection{Grid and geometry}

The simulation is performed in a spherical coordinate system $(r,\theta,\varphi)$, but we also introduce the cylindrical coordinate system $(R,Z,\varphi)$ that is useful for the analysis.

The radius ranges from $r_{in}=1$ to $r_{out}=10^5$ $\mathrm{au}$. The radial axis is discretized over 576 cells. A first patch of 512 cells follows a logarithmic progression from 1 to $10^4$ $\mathrm{au}$. The remaining cells are distributed from $10^4$ to $10^5$ $\mathrm{au}$ with a stretched spacing. The $\theta$ angle is mapped between $0$ and $\pi$ over $128$ cells. Near the poles, $32$ cells (for each side) are spread on a stretched grid, with increasing resolution towards the midplane. An additional 64 cells are used from $\theta=1.27$ to $\theta=1.87$ with uniform spacing to ensure a satisfying resolution in the equatorial region. The $\varphi$ coordinate covers the full $2\pi$ with $64$ cells evenly distributed. The total size of the computational domain is $576\times 128 \times 64$.

The configuration reaches a maximum resolution of $~10^{-2}$ $\mathrm{au}$ in the $r$ and $\theta$ directions and $~10^{-1}$ $\mathrm{au}$ in the $\varphi$ direction at $R=1 \, \mathrm{au}$, that scale linearly with the radius around the midplane. Overall, the Jeans length $\lambda_J$ is resolved by more than $20$ cells in the radial and polar direction and at least $4$ cells in the azimuthal direction.

The disc vertical extent is properly sampled with at least 10 cells per scale height $H$ at the inner boundary, where $H=\epsilon R$ is the disc geometrical scale height and assuming a canonical aspect ratio $\epsilon=0.1$. We checked that the fiducial azimuthal resolution of $64$ cells is sufficient to accurately capture non-axisymmetric perturbations by running a more resolved test, for a shorter time, with $256$ azimuthal cells. We found no qualitative difference between the two.

\subsection{Equation of state}
\label{subsec:eos}

In our setup, we do not solve the energy equation. Instead, we prescribe a barotropic equation of state (EOS) following \citet{marchand2016chemical}. As our spatial resolution is too coarse to capture the second hydrostatic core formation, this EOS reduces to:

\begin{equation}
    \label{eq:EOS_red}
    T = T_0 \sqrt{1+\left (\frac{n}{n_1}\right )^{2(\gamma-1)}}
\end{equation}

\noindent

where $n$ is the gas particle density, $T_0$ is the initial gas temperature, $\gamma=7/5$ is the adiabatic index and $n_1=10^{11} \, \mathrm{cm^{-3}}$ is the critical gas particle density.

Consequently, our effective thermal behavior could be summarized in two stages: an isothermal phase while $n<n_1$ followed by an adiabatic one. We define the formation of the first hydrostatic core as the moment where the central density reaches $n_1$. It corresponds to $t=0$ in our simulation.
   
\subsection{Non-ideal diffusivities}

The simulation takes into account Ohmic and ambipolar diffusions. To compute the associated diffusivity coefficients, we compute the steady-state abundances of the main charge carriers. For this, we use the chemical network described in Appendix~\ref{app:chemNet}. The network is solved using the code \textsc{Astrochem} \citep{maret2015astrochem} for a range of the gas densities $\rho$ and the magnetic field intensities $B$ (when relevant). The resulting diffusivities are stored in a table and, for every timestep, we read the table and perform an interpolation on-the-fly in each cell depending on the $\rho$ and $B$ value.

\subsection{Boundary conditions, internal boundaries, and restart}

The inner and outer boundary conditions are similar to a classic outflow condition, in the sense that the material can only leave the domain in the radial direction. The azimuthal magnetic field $B_\varphi$ is set to zero to prevent the angular momentum from being artificially conveyed out of the numerical domain via magnetic braking. The remaining quantities are just copied in the ghost cells from the last active one.

In the $\theta$ direction, we use an "axis" boundary condition. It is specially designed to prevent the loss of magnetic field in the polar region \citep[see appendix of][]{zhu2018global}. For the azimuthal direction, we set a classic periodic boundary condition.

For the self-gravity solver, the boundary conditions are the same as for the dynamical solver in the $\theta$ and $\varphi$ directions. In the radial direction, the gravitational potential is set to zero at the outer boundary. We define a specific "origin" inner boundary condition that expands the grid down to the center (see Appendix~\ref{app:origin}).

We implemented three internal numerical boundaries, mainly to prevent the timestep from dropping, while ensuring physical accuracy. These features include an Alfvén speed limiter, diffusivity caps \citep[following][]{xu2021formation1,xu2021formation2}, and an advection timestep limiter. A detailed discussion is provided in Appendix~\ref{app:intBdy}.

The full integration is performed following two steps. We first integrate the problem assuming a 2D axisymmetric geometry (with a single azimuthal cell) until just before the first core formation (this takes about one free-fall time). The axisymmetric assumption allows us to save computation time during the initial collapse in which the flow is quasi-axisymmetric. We then continue the integration in full 3D geometry before the first core formation, for a $100 \, \mathrm{kyr}$ integration.

Because the first step is 2D, the initial conditions for the second step are axisymmetric, which may prevent the emergence of non-axisymmetric perturbations. To alleviate this problem, we add a white noise of amplitude $\pm 0.1 \, u_\varphi$ to the azimuthal velocity when starting the 3D simulation. We checked that the angular momentum is conserved when adding this white noise.

\subsection{Initial conditions}
\label{subsec:initcond}

The initial conditions mostly follow \citet{masson2016ambipolar}. We consider a $M_0=1 \, \mathrm{M_\odot}$ spherical cloud of initial radius $r_0=2500 \, \mathrm{au}$ and uniform particle density $n_0\simeq 2\times 10^6 \, \mathrm{cm^{-3}}$. It is embedded in a $100$ times more diluted halo of radius $10^{5} \, \mathrm{au}$. The associated free-fall time is $t_{ff}=\sqrt{3\pi/32G\rho_0}\approx 22.1 \, \mathrm{kyr}$, with $\rho_0$ the initial uniform gas mass density\footnote{$\rho_0\approx 9\times 10^{-18} \, \mathrm{g \, cm^{-3}}$ .}.

The thermal over gravitational energy ratio is $\alpha=(5 r_0 c_{s0}^2)/(2 M_0 G)=0.25$, corresponding to an initial isothermal sound speed $c_{s0}\simeq 0.188 \, \mathrm{km \, s^{-1}}$. The initial temperature\footnote{We assume a mean mass per neutral particle $m_n=2.33m_p$, corresponding to the composition of the solar nebula.} is, therefore, $T_0= c_{s0}^2m_n/k_B\simeq 10 \, \mathrm{K}$.

The core is subject to solid body rotation with a ratio of rotational over gravitational energy $\beta=(\Omega_0^2 r_0^3)/(3 M_0 G)=0.02$ corresponding to a rotation rate $\Omega_0\simeq 3.9\times 10^{-13} \, \mathrm{rad \, s^{-1}}$. One difference with \citet{masson2016ambipolar} is that the background is also rotating, with a profile $\Omega(r)=\Omega_0(r/r_0)^{-2}$ for $r>r_0$ which corresponds to constant specific angular momentum along (spherical) radial lines \citep[following][]{xu2021formation1}.

Another difference is that the \emph{whole} domain is initially threaded by a uniform vertical magnetic field $B_0$ (and not only the central core). We set a mass-to-flux ratio\footnote{$\mu=\frac{M_0/(B_0 \pi r_0^2)}{(M/\phi)_{cr}}$. The core is therefore supercritical ($\mu>1$).} $\mu=2$ in unit of the critical value for collapse $(M/\phi)_{cr}=(3/2)(63G)^{-1/2}$ \citep{mouschovias1976note} which corresponds to $B_0\simeq 4\times 10^{-4} \, \mathrm{G}$.

\section{Overall evolution}
\label{sec:overall}

This section focuses on the qualitative properties of the run. First, we present the behavior of the gas and attached magnetic field during the first isothermal collapse phase and subsequent disc formation. Second, we examine the disc secular evolution properties. Third, we look at the evolution of the disc in terms of dynamics, size, and mass repartition.

\subsection{From pre-stellar collapse to disc formation}

   \begin{figure*}
   \centering
   \includegraphics[scale=1, width=\linewidth]{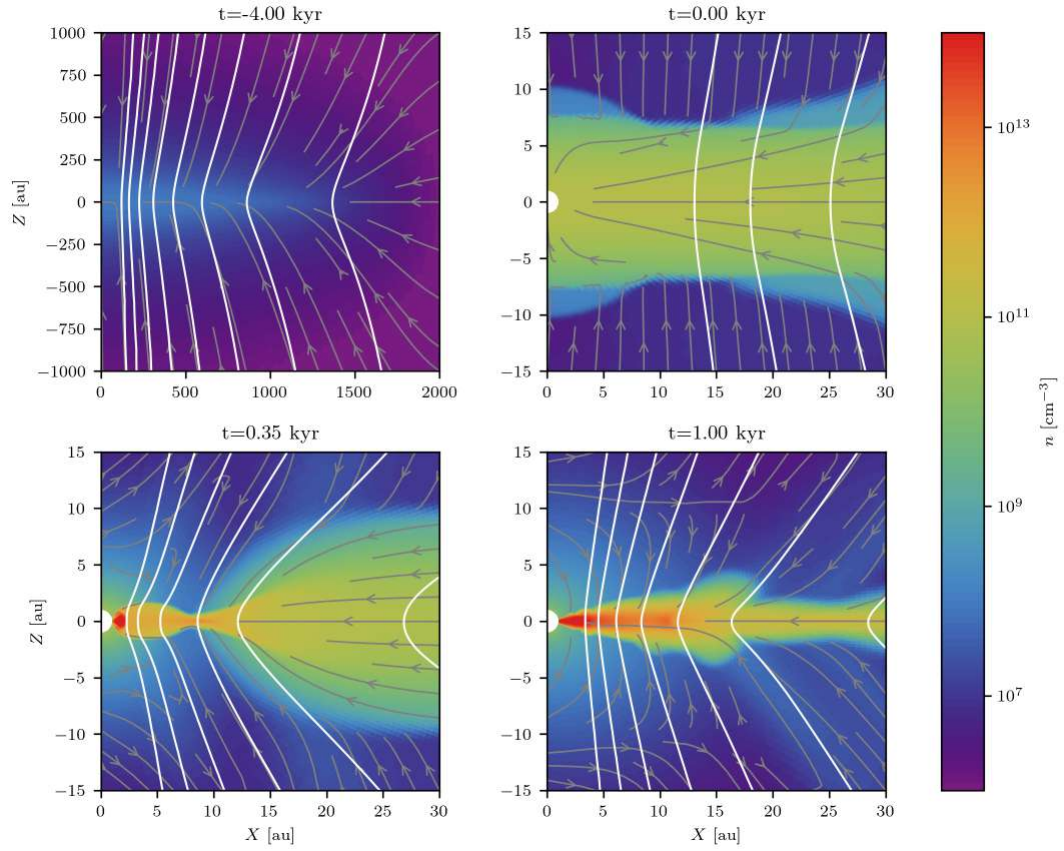}
   \caption{Snapshots of the azimuthally averaged particle density (color) with attached poloidal magnetic field lines (white contours) and poloidal velocity stream (grey arrows). From left to right: the first snapshot is a large-scale view focusing on the cloud morphology significantly before the first core formation, while the three last snapshots zoom into the first core during and after its formation.}
    \label{fig:pre_stellar_RTH}%
    \end{figure*}

We show in Fig.~\ref{fig:pre_stellar_RTH} a few snapshots of the azimuthally averaged gas particle density with attached field lines and poloidal velocity stream slightly before and just after the first hydrostatic core formation.

The first snapshot is a large-scale view of the collapsing core. It illustrates how the magnetic field acts upon the infalling material: vertically, the motion is aligned with the field and the gas is free-falling. Radially, the misalignment generates a magnetic tension that slows the collapse. The result is a flattening of the core, as well as a pinching of the magnetic field lines that are dragged along the midplane by the gas. The clear interplay between gas motions and field line deformation is a result of the low diffusivities involved at this stage of the collapse. They are inefficient at decoupling the gas and the magnetic field, which remains in a near-ideal MHD regime. 

The three last snapshots focus on what happens at a small scale, just after the first hydrostatic core formation. As the core particle density increases, it reaches the critical value $n_1$ (cf. Eq.~(\ref{eq:EOS_red})). The gas becomes adiabatic, which provides thermal support to the core that stops collapsing vertically (second snapshot). In the meantime, the radial collapse catches up and drags the magnetic field which acquires an hourglass shape. A torus-like, pressure-supported structure arises: it's the first hydrostatic core formation (third snapshot).

The large densities also increase the ambipolar and Ohmic diffusions. Inside the first core, the gas and the magnetic field are decoupled, and angular momentum accumulates. As a consequence, a small, rotationally-supported disc settles (last snapshot). The newborn disc is fed by the remnant, vertically pressure-supported gas that could never reach the radial hydrostatic equilibrium. We refer to this midplane, pressure-supported gas as the "pseudo-disc" in the following. The "envelope" denomination is more generic and  corresponds to any material that is not belonging to the disc or the seed. 

\subsection{Disc secular evolution properties}

   \begin{figure*}
   \centering
   \includegraphics[scale=1, width=\linewidth]{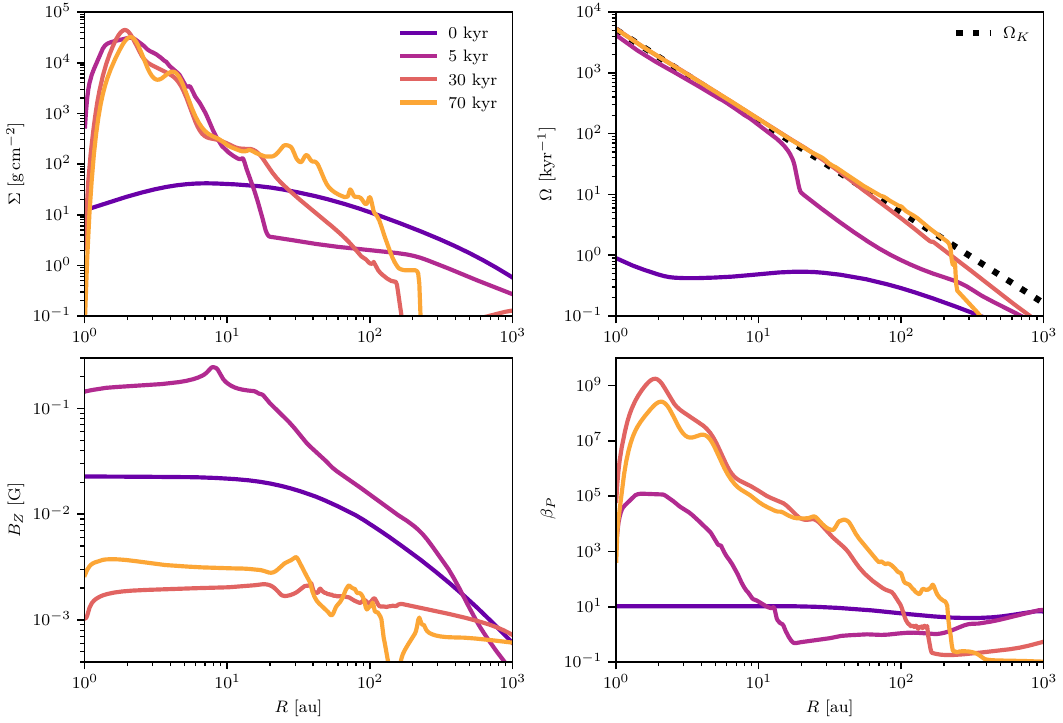}
   \caption{Azimuthally averaged surface density (top left), rotation rate (top right), poloidal magnetic field intensity (bottom left), and plasma parameter $\beta_P$ (bottom right) versus the radius, computed in the midplane. Each color corresponds to one snapshot, the darker being the disc formation epoch, while the lighter is associated with the later times of the simulation. In the top right panel, the black dashed line is the theoretical Keplerian rotation rate with $M_\star\approx 0.7 M_\odot$.}
    \label{fig:midplane_radial_prof}%
    \end{figure*}

In Fig.~\ref{fig:midplane_radial_prof} we present the azimuthally-averaged midplane surface density $\Sigma$, rotation rate $\Omega$, poloidal magnetic field intensity $B_Z$, and plasma parameter $\beta_P$ as a function of the radius, starting from the first core formation until the later times of the simulation. 

We compute the surface density as $\Sigma \equiv \int_0^{\pi} \rho r \sin \theta d\theta$, where the density is integrated along the polar angle. The density barely contributes out of the disc, which itself covers a small $\theta$ range around the midplane. Therefore, this polar integration in spherical coordinates is a convenient approximation of the vertical integration in cylindrical coordinates.

At $5 \, \mathrm{kyr}$ for $R\lesssim 10 \, \mathrm{au}$, the gas follows a steep power-law starting from a flat, high $\Sigma$. Further out, it follows a shallow power-law with low $\Sigma$. As time goes on, the steep power law becomes shallower while the transition radius increases up to $200 \, \mathrm{au}$.

The rotation rate is compared with the Keplerian prediction $\Omega_K=\sqrt{G M_\star}R^{-3/2}$ with $M_\star$ the seed mass and $R$ the cylindrical radius\footnote{$\Omega_K$ is derived taking $M_\star$ at $50 \, \mathrm{kyr}$, because $M_\star$ is roughly constant afterwards.}. At $5 \, \mathrm{kyr}$ for $R \lesssim 10 \, \mathrm{au}$, the gas is in Keplerian rotation. Further out, it is sub-Keplerian. As time goes on, the Keplerian transition radius increases up to $200 \, \mathrm{au}$. Thus, the steep, inner $\Sigma$ region is associated with rotation-supported material while the outer shallow $\Sigma$ region is sub-Keplerian.

At $5 \, \mathrm{kyr}$ for $R\lesssim 10 \, \mathrm{au}$, the poloidal magnetic field shows a plateau. Further out, it follows a power-law. The intensity of the plateau decreases with time down to a few $\mathrm{mG}$ and the plateau transition radius increases up to a few $10 \, \mathrm{au}$. Initially, the plateau is associated with the rotation-supported, steep $\Sigma$ region while the power-law is associated with the pressure-supported, shallow $\Sigma$ region.

The plateau is characteristic of the non-ideal MHD regime, responsible for decoupling the gas and $B_Z$, while the power-law indicates a near-ideal MHD regime due to the gas being less dense. A slight bump is observed at the transition radius and can be explained as follows: in the pseudo-disc region, $B_Z$ is dragged to inner radii by infalling material. Reaching the non-ideal region, most of this field cannot be conveyed any further and piles up.

Finally, the plasma parameter is defined as $\beta_P=P_{th}/P_{mag}$, where $P$ is the thermal pressure and $P_{mag}=B^2/8\pi$ is the magnetic one. Thus, $\beta_P \gg 1$ indicates a thermally-dominated gas, while $\beta_P \ll 1$ indicates a magnetically-dominated one.

At $5 \, \mathrm{kyr}$ for $R\lesssim 10 \, \mathrm{au}$, it follows a steep power-law starting from a high $\beta_P \approx 10^5$. Further out, the profile is slowly increasing from $\beta_P\approx 1$ and stays close to this limit value between the two regimes. Hence, there is a correlation between the magnetized, high surface density, rotationally-supported gas, and the thermally-dominated region.

As time goes on, the steep power law becomes shallower while the transition radius increases up to $200 \, \mathrm{au}$. The innermost region is even more thermally-dominated, reaching $\beta_P \approx 10^9$. The outer region becomes magnetically-dominated, with $\beta_P\approx 10^{-1}$. We note that for any snapshot, the limit value $\beta_P \approx 1$ is located near the transition radius in the three other profiles.

\subsection{Dynamics, size and mass repartition}

   \begin{figure*}
   \centering
   \includegraphics[scale=1, width=\linewidth]{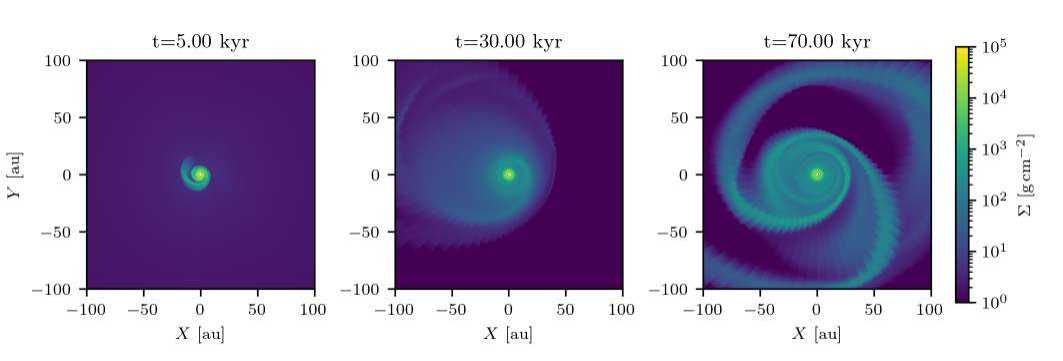}
   \caption{Gas surface density in the equatorial plane. From left to right: three characteristic snapshots illustrating the successive behaviors of the disc at $5$, $30$, and $70 \, \mathrm{kyr}$ respectively.}
    \label{fig:sigma_XY}%
    \end{figure*}
%

   \begin{figure}
   \centering
   \includegraphics[scale=1, width=0.5\linewidth]{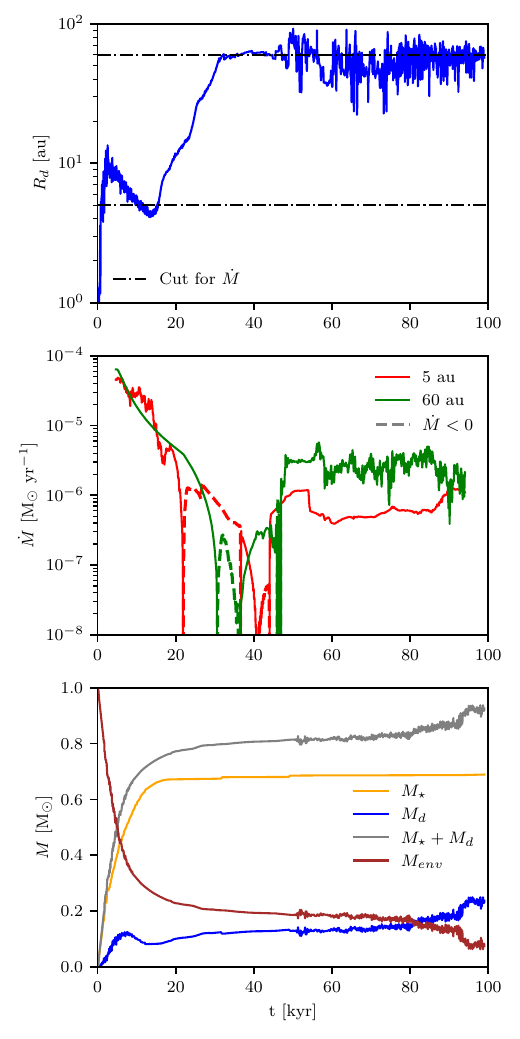}
   \caption{Top panel: radius of the disc over time. The dash-dotted black lines emphasize the radii where the mass accretion rate is inspected in the next panel. Middle panel: mass accretion rate over time near the protostar ($R=5 \, \mathrm{au}$, red) and at the maximum stable outer radius ($R=60 \, \mathrm{au}$, green). Dashed lines correspond to expanding material. Bottom panel: protostar mass (orange), disc mass (blue), the total mass of the disc-protostar system (grey), and envelope mass (brown)  over time. Data for the accretion rate are convolved in time using a $10 \, \mathrm{kyr}$ window (equivalent to $8$ orbits at $100 \, \mathrm{au}$).}
    \label{fig:mass_radius_evolution}
    \end{figure}

The disc morphology is investigated in Fig.~\ref{fig:sigma_XY}, which shows the gas surface density in the equatorial midplane at $5$, $30$, and $70 \, \mathrm{kyr}$. It covers the different dynamical states experienced by the disc during the secular integration: first, the disc is small and subject to spiral density waves. Second, it smoothes out and builds up, apart from one large, streamer-like spiral arm. Finally, the outer disc triggers new spirals that propagate to the inner radii.

In the middle panel, the disc exhibits a slightly eccentric morphology. We caution that this is probably a consequence of the fixed point mass assumption. In principle, while accreting mass with linear momentum, the point mass should move. Because this motion is not taken into account, there is a non-zero indirect term from gravity which makes the disc eccentric. That being said, we think it does not significantly affect the results, and the eccentricity disappears afterwards (see right panel).

Figure \ref{fig:mass_radius_evolution} gives an overview of the evolution of the disc radius ($R_\mathrm{d}$), the accretion rate ($\dot{M}$) and the mass repartition between the seed ($M_\star$), the disc ($M_\mathrm{d}$) and the envelope ($M_\mathrm{env}$) over the $100 \, \mathrm{kyr}$ of the simulation.

The top panel follows the evolution of the disc radius $R_\mathrm{d}$. To compute it, we assume that cells satisfying $u_\varphi \geq 0.9 V_{K}$, $u_\varphi > u_{R}$ and $u_\varphi > c_s$ are part of the disc ($V_{K}$ and $c_s$ are the local Keplerian and sound velocities). The disc radius is then defined as the azimuthal median of the outermost radii that satisfy this criterion in the midplane.

As the disc forms after a few $\mathrm{kyr}$, its radius reaches $10 \, \mathrm{au}$ and immediately starts decreasing until $15 \, \mathrm{kyr}$. Then, it increases smoothly up to $60 \, \mathrm{au}$ after $30 \, \mathrm{kyr}$ where it remains constant until $45 \, \mathrm{kyr}$. After this point, the radius is subject to chaotic fluctuations around $60 \, \mathrm{au}$ and remains so until the end of the simulation.

The middle panel follows the evolution of $\dot{M}$ near the protostar ($R=5 \, \mathrm{au}$) and at the maximum stable outer radius ($R=60 \, \mathrm{au}$). We perform an azimuthal integration over $2\pi$ and a vertical integration between $\pm H$. In the following, H always refers to the disc geometrical scale height. Data are convolved in time using a $10 \, \mathrm{kyr}$ window (equivalent to $8$ orbits at $100 \, \mathrm{au}$). We caution that, in doing so, we smooth out short-timescale events occurring in the disc to focus on secular events. For instance, the fact that the accretion rate oscillates on the spiral dynamical timescale \citep{tomida2017grand} is verified, but hidden due to the large smoothing window.

We first focus on $\dot{M}(R=5 \, \mathrm{au})$, which gives a good proxy for the accretion onto the protostar. As the disc forms, the protostar accretes at a strong rate of a few $10^{-5} \, \Macc$ that immediately starts decreasing until $20 \, \mathrm{kyr}$. There, it reverses with a negative rate around $10^{-6} \, \Macc$, which means that gas is expanding consistently with $R_\mathrm{d}$ and the protostar is no more accreting. After $35 \, \mathrm{kyr}$, the expansion episode stops and standard protostar accretion is back. It is first small, with strong variations between $10^{-8}$ and a few $10^{-7} \, \Macc$. After $45 \, \mathrm{kyr}$ the range increases and stabilizes between a few $10^{-7}$ and $10^{-6} \, \Macc$.

For $\dot{M}(R=60 \, \mathrm{au})$, the material is not part of the disc until $30 \, \mathrm{kyr}$ and only marginally belongs to the disc afterwards due to radius variability, such that we essentially probe the pseudo-disc accretion. As the disc forms, the pseudo-disc accretes at a strong rate of a few $10^{-5} \, \Macc$ that immediately start decreasing until $30 \, \mathrm{kyr}$. There, it reverses with a negative rate of a few $10^{-7} \, \Macc$. It is shortly and slightly expanding because $R_\mathrm{d}$ stops growing there. After $35 \, \mathrm{kyr}$, the expansion episode stops and standard pseudo-disc accretion is back. Note that the $\dot{M}(R=60 \, \mathrm{au})$ is about one order of magnitude larger than $\dot{M}(R=5 \, \mathrm{au})$, indicating that the disc density structure is still evolving and that proper steady-state has not yet been reached.

Finally, the bottom panel shows the evolution of the mass repartition between the seed, the disc, and the envelope. $M_\star$ accounts for any mass falling below $R_{in}$. $M_\mathrm{d}$ is computed by summing $\rho dV$ over any cell matching the disc criterion. The envelope mass is what is left of the initial $1 \, \mathrm{M_\odot}$ cloud.

As the disc forms, $M_\star$ grows to $0.7 \, \mathrm{M_\odot}$ until $15 \, \mathrm{kyr}$ and stagnates afterwards. In the meantime, $M_\mathrm{d}$ reaches a maximum $0.15 \, \mathrm{M_\odot}$ and immediately starts decreasing until $15 \, \mathrm{kyr}$. Then, it increases smoothly to $0.15 \, \mathrm{M_\odot}$ after $30 \, \mathrm{kyr}$ where it remains constant until $45 \, \mathrm{kyr}$. After this point, the disc mass keeps increasing while oscillating and remains so until the end of the simulation. The final disc mass is $0.25 \, \mathrm{M_\odot}$.

$M_\mathrm{env}$ is rapidly decreasing until $10 \, \mathrm{kyr}$. Most of the lost mass ends up in the seed, the rest becomes part of the disc. After $5 \, \mathrm{kyr}$, the envelope mass becomes negligible compared to $M_\star$, and the decaying slope is shallower. The envelope is mainly accreted onto the disc. After $80 \, \mathrm{kyr}$, $M_\mathrm{env}$ becomes negligible compared to $M_\mathrm{d}$.

\section{Accretion history}
\label{sec:fidAccretion}

In this section, we study the accretion history of the disc based on the evolution of key physical quantities (surface density, magnetic field...). We isolate the main driving mechanisms for accretion, address their relevance throughout the disc evolution, and derive a comprehensive scenario over three accretion phases.

\subsection{Driving accretion mechanisms}
\label{subsec:drivingMech}

   \begin{figure}
   \centering
   \includegraphics[scale=1, width=0.5\linewidth]{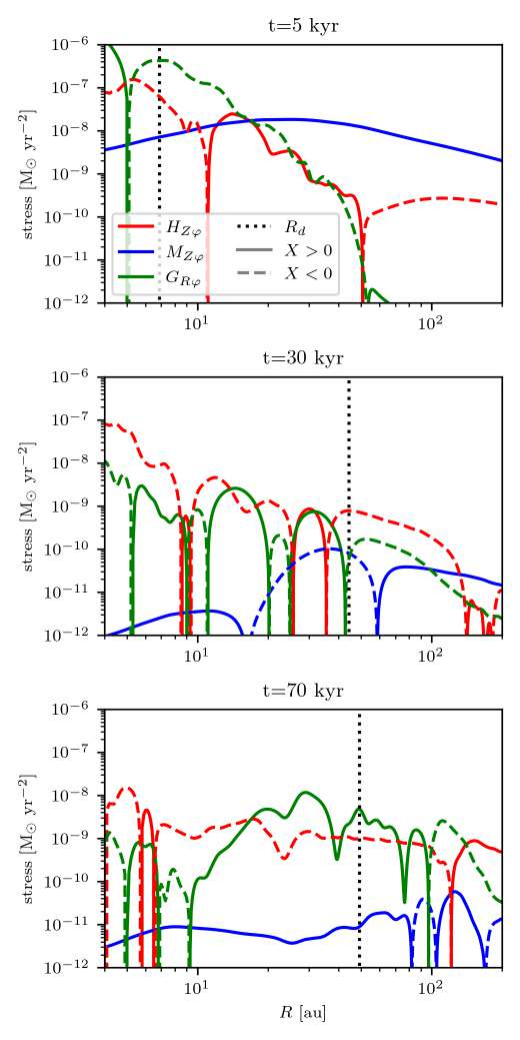}
   \caption{From top to bottom we focus on $t=5$, $30$ and $70 \, \mathrm{kyr}$ respectively. Each panel presents $G_{R\varphi}$ (green), $M_{Z\varphi}$ (blue) and $H_{Z\varphi}$ (red) versus the radius. Solid and dashed lines are associated with positive and negative stresses respectively. The black dotted line is the corresponding disc radius. Data are convolved in time using a $10 \, \mathrm{kyr}$ window (equivalent to $8$ orbits at $100 \, \mathrm{au}$).}
    \label{fig:stresses_over_phases}%
    \end{figure}
%

   \begin{figure}
   \centering
   \includegraphics[scale=1, width=\linewidth]{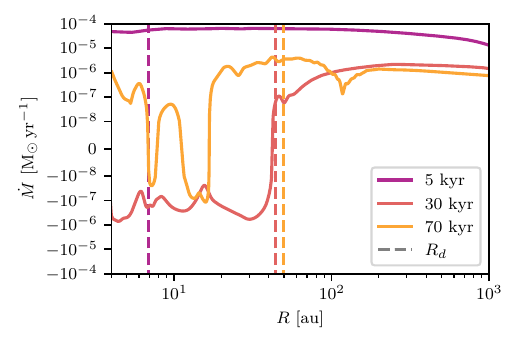}
   \caption{Mass accretion rate versus the radius. Each color corresponds to one snaphsot at $5$, $30$ and $70 \, \mathrm{kyr}$ respectively. The lighter, the later. Dashed lines represent the disc radius associated with each epoch. Data are convolved in time using a $10 \, \mathrm{kyr}$ window (equivalent to $8$ orbits at $100 \, \mathrm{au}$).}
    \label{fig:Macc_vs_R_over_phases}%
    \end{figure}

Protoplanetary discs are rotationally supported structures. In this context, accretion is only possible if there are one or several mechanisms capable of extracting angular momentum from the gas.

To properly account for each of these mechanisms, we derive the conservation of angular momentum in the cylindrical coordinates system $(R, Z,\varphi)$ for the case of a self-gravitating, magnetized rotating disc \citep[adapted from][]{lesur2021magnetohydrodynamics}:

\begin{equation}      \label{eq:momCons}
    \overline{\rho u_R}\partial_R\left (\Omega R^2\right )+\frac{1}{R}\partial_R\left ( R^2 \underbrace{\overline{\Pi_{R\varphi}}}_{\text{radial stress}} \right )+R\underbrace{\left [ \langle \Pi_{Z\varphi}  \rangle \right ]_{-H}^{H}}_{\text{surface stress}}=0
\end{equation}

\noindent
with

\begin{align}
    \label{eq:Rstress}&\Pi_{R\varphi} = \rho u_R V_\varphi + \frac{g_R g_\varphi}{4\pi G} - \frac{B_R B_\varphi}{4\pi} \\
    &\label{eq:Zstress}\Pi_{Z\varphi} = \rho u_Z V_\varphi + \frac{g_Z g_\varphi}{4\pi G} - \frac{B_Z B_\varphi}{4\pi}
\end{align}

\noindent
where $V_\varphi=u_\varphi-\frac{1}{2H}\overline{u_\varphi}$, $\textbf{g}$ is the gravitational field and

   \begin{equation}
   \label{eq:defAvg}
   \begin{array}{ccc}
        \langle X \rangle = \frac{1}{2\pi} \int_0^{2\pi} X d\varphi & \text{ and } &
        \overline{X} = \int_{-H}^{H} \langle X \rangle dZ
   \end{array}
   \end{equation}

\noindent
for any quantity $X$ and $[X]_{-H}^H=X(Z=H)-X(Z=-H)$.

There are therefore six mechanisms acting upon the angular momentum transport in our case: hydrodynamical transport (first term in Eqs.~(\ref{eq:Rstress})-(\ref{eq:Zstress})), gravitational transport (second term) and magnetic transport (third term), each of them generating both a radial stress (Eq.~(\ref{eq:Rstress})) and a surface stress (Eq.~(\ref{eq:Zstress})).

Among these quantities, we want to focus on the three main levers identified in previous works \citep{xu2021formation1,xu2021formation2} as the preponderant mechanisms at stake for such massive, embedded and magnetized discs: the radial gravitational stress $G_{R\varphi}$, the magnetic braking $M_{Z\varphi}$ (corresponding to the surface magnetic stress), and the surface hydrodynamical stress $H_{Z\varphi}$. Each of them is labeled as follows:

\begin{align}
    \label{eq:GRstress}
    &G_{R\varphi}\equiv \frac{1}{R}\partial_R \left [R^2 \frac{\overline{g_R g_\varphi}}{4\pi G} \right]\\
    \label{eq:MZstress}
    &M_{Z\varphi}\equiv -R\left [\frac{\langle B_Z B_\varphi \rangle}{4\pi} \right ]_{-H}^H\\
    \label{eq:KZstress}
    &H_{Z\varphi}\equiv R\left [\langle \rho u_Z V_\varphi \rangle \right ]_{-H}^H
\end{align}

Figure \ref{fig:stresses_over_phases} presents a comparison of each stress for three snapshots (with time increasing from top to bottom). A positive torque is associated with the extraction of angular momentum from the disc while a negative torque brings angular momentum to the disc.

At $5 \, \mathrm{kyr}$, the gravitational stress is positive and dominant in the disc. It transports angular momentum from the inside out. In the meantime, magnetic braking is positive and dominant in the pseudo-disc. It extracts angular momentum from the gas. The hydrodynamical stress can be significant but is never dominant in the innermost $200 \, \mathrm{au}$.

At $30 \, \mathrm{kyr}$, the hydrodynamical stress is negative and dominant in the disc and inner pseudo-disc. It brings angular momentum from the envelope. In the meantime, magnetic braking is positive and dominant in the outer pseudo-disc. Yet, its intensity has significantly decreased. The gravitational stress can be significant but is never dominant in the innermost $200 \, \mathrm{au}$.

At $70 \, \mathrm{kyr}$, the hydrodynamical stress is negative and dominant in the inner disc. In the meantime, the gravitational stress is positive and dominant in the outer disc and the pseudo-disc. The magnetic braking is essentially positive but never significant in the innermost $200 \, \mathrm{au}$.

The relative importance of each stress can be connected with Fig.~\ref{fig:Macc_vs_R_over_phases}, which shows the accretion rate versus the radius for the same snapshots. $\dot{M}$ is computed as in the middle panel of Fig.~\ref{fig:mass_radius_evolution}. A positive rate corresponds to accretion while a negative one is associated with expansion.

At $5 \, \mathrm{kyr}$, both the disc and the pseudo-disc efficiently accrete at a few $10^{-5} \, \Macc$. At $30 \, \mathrm{kyr}$, the disc expands with $\dot{M}$ ranging between $-10^{-7}$ and $-10^{-6} \, \Macc$ and the pseudo-disc accretes at $10^{-6} \, \Macc$. The disc is therefore growing because of inside-out expansion and accumulation at the edge. At $70 \, \mathrm{kyr}$, the inner disc has no clear trend. It switches between accretion and expansion. In the meantime, the outer disc and the pseudo-disc accrete at $\dot{M}\sim 10^{-6} \, \Macc$.

Accretion therefore results from the relative importance of each stress in the disc and the pseudo-disc. Understanding the secular evolution of these stresses is the key to understanding the different accretion behaviors.

\subsection{Secular accretion scenario}
\label{subsec:secular_scenario}
   \begin{figure*}
   \centering
   \includegraphics[scale=1, width=\linewidth]{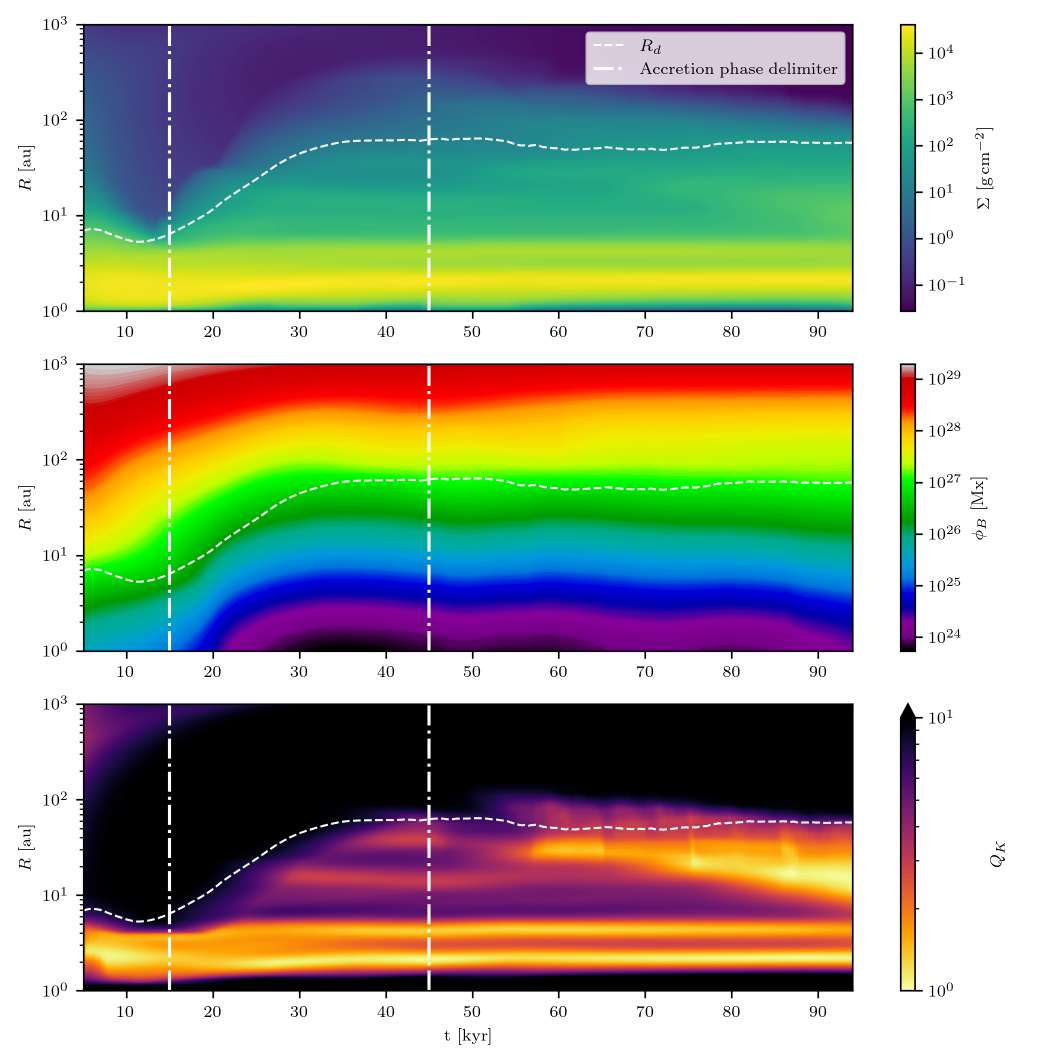}
   \caption{Spacetime diagrams of the surface density (top panel), poloidal magnetic flux (middle panel), and Toomre parameter (bottom panel). Colors and radii are in log scale. The dashed white line corresponds to the disc radius, and dash-dotted white lines delimit the three accretion phases. Data were convolved in time using a $10 \, \mathrm{kyr}$ window (equivalent to $8$ orbits at $100 \, \mathrm{au}$).}
    \label{fig:sigmaFluxToomre}%
    \end{figure*}
%

   \begin{figure*}
   \centering
   \includegraphics[scale=1, width=\linewidth]{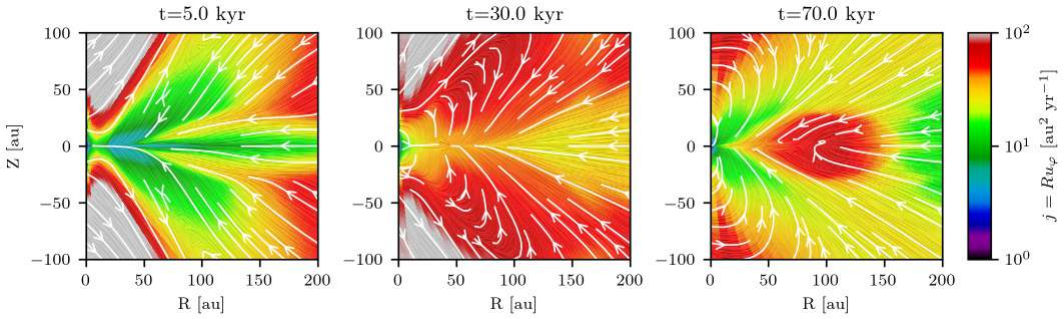}
   \caption{From left to right we focus on $t=5$, $30$, and $70 \, \mathrm{kyr}$ respectively, corresponding to each accretion phase. Each panel presents the specific angular momentum $j$ in colors, with the attached poloidal velocity flow $\vec{V_p}$ as white quivers.}
    \label{fig:jspec_over_phases}%
    \end{figure*}

In the following, we derive the disc accretion scenario in three phases by looking into the physical quantities underlying each stress.

First, spiral density waves are known to be an efficient angular momentum carrier. In our simulation, they are observed simultaneously with high accretion rates in the disc, making them a good candidate to explain accretion. Typically, Gravitational Instabilities (GI) are triggered when $M_{d}\gtrsim\epsilon M_\star$ \citep[][Eq. 12]{armitage2011dynamics}. The aspect ratio of the disc is $\epsilon\lesssim 0.1$ during the simulation, while $M_\mathrm{d}>0.1M_\star$. Thus, our disc lies in the adequate regime, indicating that our spiral density waves are probably triggered by GI.

In order to characterize the stability of the disc with respect to its own gravity, we use a simplified version of the Toomre parameter $Q$ \citep{toomre1964gravitational,goldreich1965gravitational,goldreich1965ii}. Assuming that the gas is in Keplerian rotation, the simplified Toomre parameter $Q_K$ writes as \citep{xu2021formation1,xu2021formation2}:

\begin{equation}      \label{eq:myToomre}
    Q_K = \left ( \frac{c_s \Omega_K}{\pi G  \Sigma}\right )_{Z=0}
\end{equation}

Many studies discuss the critical value for stability, which depends on the assumptions on the disc thickness or the perturbations linearity \citep{toomre1964gravitational,goldreich1965gravitational,goldreich1965ii,gammie2001nonlinear,wang2010equilibrium}. Based on these works, we expect the disc to be unstable when $Q_K\sim 1$ while keeping in mind that the lower $Q_K$, the more likely the GI.

Second, the magnetic braking is a function of the poloidal magnetic field. Hence, the braking efficiency is controlled by the amount of poloidal magnetic flux stored in the gas. It is computed from the magnetic vector potential $\mathbf{A}$:
\begin{equation}
\phi_B = R\langle A_\varphi(Z=0)\rangle
\end{equation}

To complete our diagnostics, the hydrodynamical vertical stress is a function of the specific angular momentum $j=Ru_\varphi$ transported by the poloidal velocity flow $\vec{V_p} = u_R\vec{e_R}+u_Z\vec{e_Z}$.

We present in Fig.~\ref{fig:sigmaFluxToomre} a series of spacetime diagrams connecting the surface density of the gas (top panel), the poloidal magnetic flux (middle panel), and the Toomre parameter (bottom panel). Fig~\ref{fig:jspec_over_phases} shows the specific angular momentum with the attached poloidal flow.  We discuss below the different phases that we find from these figures.

\subsubsection{Phase I: a small, GI-driven disc}

From the top panel of Fig.~\ref{fig:sigmaFluxToomre}, we see that until $15 \, \mathrm{kyr}$, a high surface density region concentrates to the innermost $10 \, \mathrm{au}$ corresponding to the disc. At the transition, there is a sharp drop in density, and the pseudo-disc is left with a low surface density. In parallel, the middle panel shows that a lot of magnetic flux is stored in the gas, especially within the pseudo-disc region, while the bottom panel presents a Toomre parameter that is of the order of unity in the disc region, which is, therefore, GI unstable.

In addition, the left panel of Fig.~\ref{fig:jspec_over_phases} shows that the pseudo-disc provides material to the disc that has low specific angular momentum. At intermediate altitudes, the envelope provides material with higher $j$ but does not reach the inner regions corresponding to the disc. At high altitudes, it provides material to the disc through its surfaces, but with low specific angular momentum.

Thus, in the first phase, the disc feeds the protostar thanks to GI-triggered spiral density waves that efficiently remove angular momentum at low radii. The instability is itself sustained by the mass influx from the pseudo-disc, driven by a powerful magnetic braking. There is a specific angular momentum contribution from the pseudo-disc/envelope to the disc, but it is not significant.

\subsubsection{Phase II: disc expansion fed by the envelope}
\label{sssect:phase2}
From the top panel of Fig.~\ref{fig:sigmaFluxToomre}, we see that between $15$ and $45 \, \mathrm{kyr}$, the surface density increases at $R\gtrsim 10 \, \mathrm{au}$, while the disc grows from inside out. This is synchronous with a continuous outwards advection of the magnetic flux in the middle panel. We emphasize that at $10 \, \mathrm{au}$, $\phi_B$ decreases by roughly $2$ orders of magnitude in the $20$ first $\mathrm{kyr}$. In the meantime, from the bottom panel, the Toomre parameter exhibits quite complex behavior. For $R \lesssim 10 \, \mathrm{au}$, $Q_K$ is close to unity. This trend, along with the persistence of two low $Q_K$ rings until the end of the simulation, predicts that the inner disc should trigger spiral density waves which we do not observe for these times and for these radii. Instead, two persistent ring-like features are observed in the surface density. Such rings are ubiquitous in self-gravitating disc simulations \citep{durisen2005hybrid,boley2006thermal,michael2012convergence,steiman2013convergence,steiman2023mass} and are believed to be a common product of GI. 
Between $10$ and $60 \, \mathrm{au}$, a significant decrease down to $Q_K \sim 5$ is observed, as a response to the surface density increase, but this is not enough to enter a GI regime. The disc therefore smoothes out.

In addition, the middle panel of Fig.~\ref{fig:jspec_over_phases} shows that the pseudo-disc has become more efficient at providing angular momentum to the disc. At intermediate and high altitudes, the envelope now provides material with a large specific angular momentum, among which a significant part reaches the disc.

Thus, in the second phase, the disc stabilizes and smoothes out. It can't accrete, since no angular momentum transport mechanism is efficient enough in this phase. On the contrary, it gains angular momentum from the envelope. The net result is a radial expansion of the disc, and the surface density power law becomes shallower. In the process, the accumulated magnetic flux is advected outwards in a flux-freezing manner, hence a decrease in the magnetisation. This is discussed in Sect.~\ref{ssect:fluxfreeze}.

\subsubsection{Phase III: GI-driven outer disc}

From the top panel of Fig.~\ref{fig:sigmaFluxToomre}, we see that the disc radius expansion is halted around $45 \, \mathrm{kyr}$ and stabilizes at $60 \, \mathrm{au}$ until the end of the simulation. In the meantime, the surface density front stops around $100 \, \mathrm{au}$, with a slight tendency to decrease by the end of the simulation. The flux outwards advection in the middle panel stops as well, and there is even some late inwards advection in the disc. In the bottom panel, at roughly $55 \, \mathrm{kyr}$, $Q_K$ becomes of the order of unity in the outermost part of the disc. The low value propagates to inner radii afterwards, until the end of the simulation.

In addition, the right panel of Fig.~\ref{fig:jspec_over_phases} shows that the pseudo-disc starts to accrete low $j$ material again. At intermediate altitudes, the specific angular momentum is the strongest, near the outer disc surfaces. From high altitudes, the envelope still provides material with significant $j$ to the inner disc.

Thus, in the third phase, the disc does not receive enough angular momentum to keep expanding. The pseudo-disc is still braked and accretes onto the disc. The surface density profile steepens at the disc edge and triggers a new GI producing new spirals that propagate to inner radii. Hence the outer disc can accrete. This final state holds for the remaining $55 \, \mathrm{kyr}$ suggesting that the disc ends up in a GI-regulated state. We explore this idea in the following.

\subsection{Flux-freezing during disc expansion}
\label{ssect:fluxfreeze}

   \begin{figure}
   \centering
   \includegraphics[scale=1, width=\linewidth]{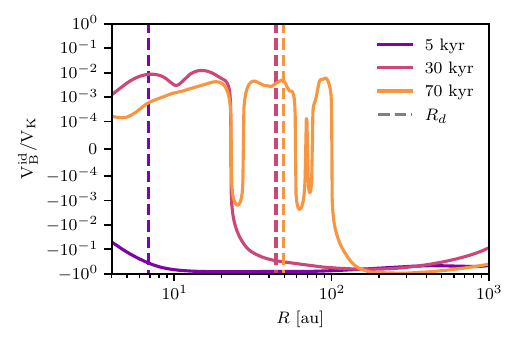}
   \caption{Ideal magnetic flux velocity transport versus the radius. Each color corresponds to one snapshot at $5$, $30$, and $70\, \mathrm{kyr}$ respectively. The lighter, the later. Dashed lines represent the disc radius associated with each epoch. Data are convolved in time using a $10\, \mathrm{kyr}$ window (equivalent to $8$ orbits at $100\, \mathrm{au}$).}
    \label{fig:fluxTransport}%
    \end{figure}

In Sect.~\ref{sssect:phase2}, we find that while the disc is expanding, the magnetic flux is advected along with the gas consistently with a flux-freezing behavior. Here we discuss whether advection is indeed the dominant contribution to the magnetic field transport during the disc secular evolution. We define the midplane, azimuthally-averaged, ideal, poloidal magnetic flux velocity transport $V_B^{id}$ as \citep[adapted from][]{lesur2021systematic}:

\begin{equation}
    \label{eq:idealVB}
    V_B^{id} = \frac{\langle \mathcal{E}^{id}_\varphi (Z=0) \rangle}{\langle B_Z (Z=0)\rangle}
\end{equation}

where $\mathcal{E}_\varphi^{id}=u_R B_Z$ comes from the first term of Eq.~(\ref{eq:ohm}). Physically, it corresponds to the magnetic flux variation associated to advection.

Figure~\ref{fig:fluxTransport} presents $V_B^{id}$ versus the radius for each phase, along with the location of the disc radius. It is normalized by the Keplerian velocity $V_K$. A positive velocity is associated with outwards transport while a negative velocity is associated with inwards transport. In the following, we compare the prediction for the flux transport from advection only with the actual evolution from Fig.~\ref{fig:sigmaFluxToomre}, middle panel. If they match, we conclude that the advection is mainly responsible for the flux transport, i.e the gas exhibits a flux-freezing behavior.

At $5\, \mathrm{kyr}$, the ideal transport predicts strong ($\approx -0.1$) inwards advection for the flux inside the disc, while we see it starts spreading in Fig.~\ref{fig:sigmaFluxToomre} (middle panel, phase I). Thus, diffusive transport dominates over advection. At $30\, \mathrm{kyr}$, the ideal transport predicts significant ($\approx 0.01$) outwards advection for the flux inside the disc, which is consistent with the flux recession in Fig.~\ref{fig:sigmaFluxToomre} (middle panel, phase II). Thus, the flux is advected by the spreading disc. At $70\, \mathrm{kyr}$, the ideal transport predicts low ($<0.001$) outwards advection in the inner smooth disc and a slightly stronger ($>0.001$) outwards advection in the outer spiral-driven disc, while the flux seems slightly advected back inwards on the long term in Fig.~\ref{fig:sigmaFluxToomre} (middle panel, phase III). This discrepancy suggests that accretion is not the main driver of flux transport in this phase.

Thus, we conclude that diffusion is responsible for flux leaking essentially during phase I and III. Conversely, during phase II, advection overcomes diffusion and the field is just diluted in the expanding disc.

\subsection{A GI-regulated final secular evolution}
\label{subsec:gi-reg-evol}
   \begin{figure}
   \centering
   \includegraphics[scale=1, width=\linewidth]{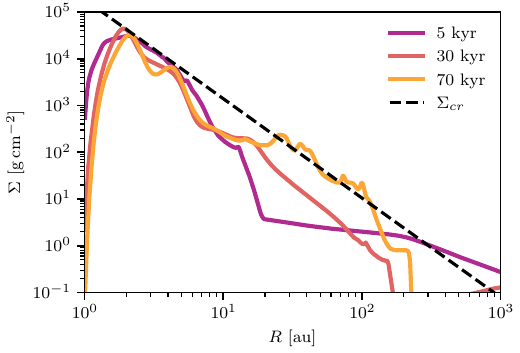}
   \caption{Disc surface density versus radius. Each color focuses on one accretion phase at $5$, $30$, and $70 \, \mathrm{kyr}$ respectively. The lighter, the later. The black dashed line is the predicted critical surface density from Eq.~(\ref{eq:sigmaCr}).}
    \label{fig:sigma_vs_cr}%
    \end{figure}

In Sect.~\ref{subsec:secular_scenario}, we conclude that the disc angular momentum process is dominated by GI-driven spiral density waves when accreting. Self-gravitating discs are prone to enter a marginally unstable, self-regulated gravito-turbulent state where the Toomre parameter is maintained around the critical value $Q \sim 1$ \citep{gammie2001nonlinear}.

From Eq.~(\ref{eq:myToomre}), $Q_K$ is controlled by two main levers : surface density $\Sigma$ and temperature $T$ (through $c_s=\sqrt{k_B T/m_n}$). Yet, we use a barotropic EOS, such that $T$ is set by the density. In this specific case, \citet{xu2021formation1} argues that the disc stability is controlled by the surface density alone. It can enter GI through $\Sigma$ increase, as a result of mass influx from the environment. Once unstable, such a disc can be brought back into stability only by lowering $\Sigma$, if authorized to spread or to significantly accrete.

To probe this phenomenon, we derive a critical surface density $\Sigma_{cr}$ as a function of the radius, above which the disc should become unstable :

\begin{equation}
    \label{eq:sigmaCr}
    \Sigma_{cr}(R) = \Sigma_0 R^{(\gamma+2)/(\gamma-3)} = \Sigma_0 \left (\frac{R}{1 \, \mathrm{au}}\right)^{-2.125}
\end{equation}

\noindent
where

\begin{equation}
    \label{eq:sigma0}
    \Sigma_0=\left [\frac{c_{s,0}}{\pi}\sqrt{\frac{M_\star}{G}}(\epsilon \rho_1)^{(1-\gamma)/2} \right]^{2/(3-\gamma)}\approx 2\times 10^{5} \, \mathrm{g \, cm^{-2}}
\end{equation}

\noindent
with $\rho_1=n_1\cdot m_n$ the critical mass density for which the gas becomes adiabatic.

$\Sigma_{cr}$ is calculated from Eq.~(\ref{eq:myToomre}) with the following assumptions :

\begin{itemize}
    \item $Q_K=1$.
    \item $c_s = c_{s,0}\left (\frac{\rho}{\rho_1}\right )^{(\gamma-1)/2}$ (adiabatic regime).
    \item $\rho=\frac{\Sigma}{H}=\frac{\Sigma}{\epsilon R}$.
\end{itemize}

Most of the data used for the calculation are detailed in Sects.~\ref{subsec:eos} and \ref{subsec:initcond}, and we take $\gamma=1.4$, $\epsilon=0.1$ and $M_\star=0.7 \, M_{\odot}$.

The critical surface density is reported as a dashed black line in Fig.~\ref{fig:sigma_vs_cr}, along with the measured disc surface density for three snapshots spanning over each phase.

At $5 \, \mathrm{kyr}$, the steep power-law at $R\lesssim 10 \, \mathrm{au}$ matches the critical value between $2$ and $4 \, \mathrm{au}$, and stands below further out. This is consistent with the spiral-driven disc in phase I. At $30 \, \mathrm{kyr}$, the power-law spreads, while staying below $\Sigma_{cr}$. This is consistent with the smooth discs in phase II. At $70 \, \mathrm{kyr}$, the inner disc stands below the critical surface density while the outer disc saturates at $\Sigma_{cr}$, around which it oscillates. This is consistent with the spiral-driven outer disc in phase III.

Hence, any deviation from the stability regime enforced by self-gravity leads to negative feedback that promotes accretion. In this sense, the disc is shaped by self-gravity. In the case where a sustained influx of material increases locally the surface density, the disc enters a self-regulated state where $R_\mathrm{d}$ stabilizes around $60 \, \mathrm{au}$. This emphasizes the role of the environment interacting with the disc.

\section{Interaction with the parent molecular cloud : a large-scale accretion streamer}
\label{sec:streamer}

The most striking evidence of an interaction between the molecular cloud and the protostar-disc system is the appearance of a large-scale, streamer-like spiral arm driving asymmetric infall between the remnant cloud and the central protostar-disc system. Here, we discuss the accretion streamer morphology and kinematics and investigate on how it connects to the central system. We also compute some observational properties.

\subsection{Streamer spatial and velocity distribution : a sheet-like morphology}

   \begin{figure*}
   \centering
   \includegraphics[scale=1, width=\linewidth]{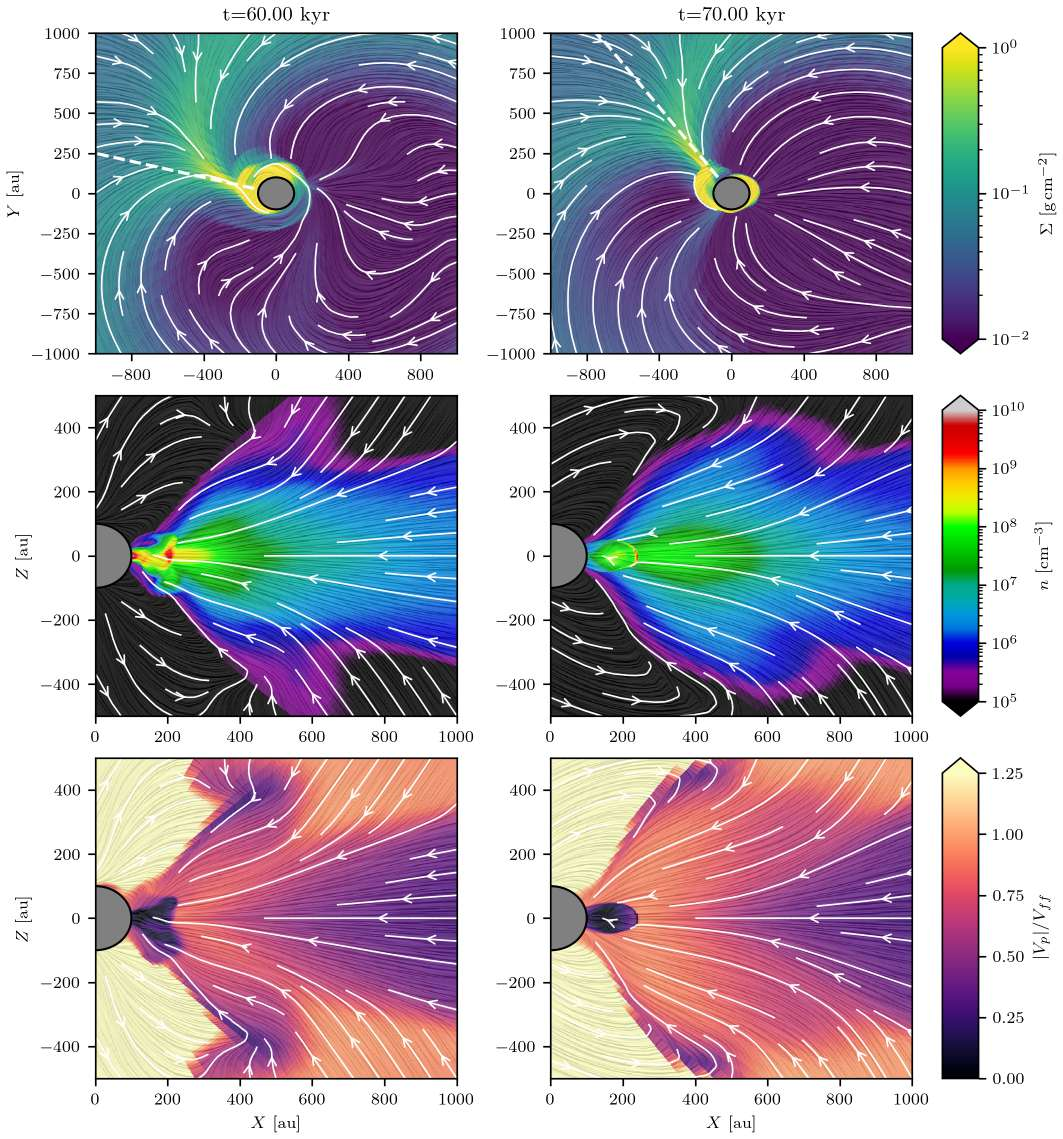}
   \caption{Top row: large-scale equatorial slice of the gas surface density with associated equatorial velocity flow (white quivers). A dashed white line indicates the azimuth of the main streamer, for which poloidal slices are performed. Middle row: large-scale poloidal slice of the gas particle density with associated poloidal velocity flow (white quivers). Bottom row: large-scale poloidal slice of the gas poloidal velocity, normalized by the free-fall velocity, with associated poloidal velocity flow (white quivers). Each column corresponds to a late time snapshot, respectively at $60$ (left column) and $70 \, \mathrm{kyr}$ (right column). In each plot, a line-integrated convolution treatment is applied to emphasize the gas streamlines. The innermost $100 \, \mathrm{au}$ are masked by a grey circle to allow for adequate contrast.}
    \label{fig:streamer}%
    \end{figure*}
%

   \begin{figure}
   \centering
   \includegraphics[scale=1, width=\linewidth]{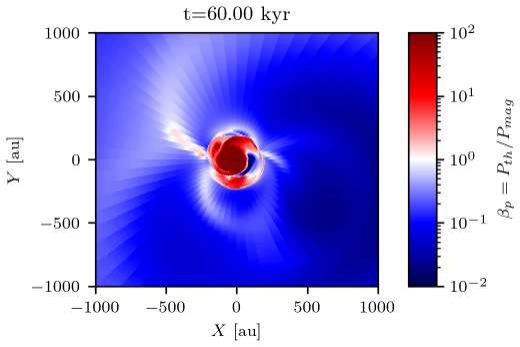}
   \caption{Large-scale equatorial slice at $t=60 \, \mathrm{kyr}$. Colors are the plasma parameter $\beta_P$.}
    \label{fig:beta_streamer}%
    \end{figure}

Figure \ref{fig:streamer} presents a large-scale equatorial slice of the gas surface density (top row) along with poloidal slices of the gas particle density (middle row) and poloidal velocity (bottom row) performed at the azimuth of the streamer for two late time snapshots. The poloidal velocity is normalized by the free-fall velocity $V_{ff}=\sqrt{2}V_K$.

Focusing on the top row, we see that the environment of the protostar-disc system is divided between a low surface density "bubble" and an extended region of growing surface density that culminates in an azimuthally localized channel of gas at lower radii. This is the main accretion streamer, towards which the equatorial velocity flow is converging. The streamer structure extends up to approximately $1000 \, \mathrm{au}$ and connects to the disc. A second converging flow is observed in the low-density "bubble", corresponding to an additional fainter streamer. The streamers rotate between the two snapshots.

The middle row shows that the gas particle density is vertically stratified, with densities ranging between $10^6 \, \mathrm{cm^{-3}}$ close to the polar axis and $10^8 \, cm^{-3}$ in the midplane. Overall, the streamer converges in a sheet-like morphology that is even better identified when represented in three dimensions \footnote{A 3D model of the streamer at $70 \, \mathrm{kyr}$ based on particle density contours is available on 
\url{https://sketchfab.com/3d-models/streamer-1407-fid-66107da9c4854078aadac140de9f4e73}.}.

Finally, the bottom row illustrates two different dynamical behaviors. Far from the midplane, the flow is near free-fall. The gas is channeled towards lower radii at constant velocity and falls directly above (and below) the disc. Near the midplane, the large-scale velocity is smaller. A gradient is observed towards lower radii as the gas catches up with more elevated material, reaching near free-fall velocity. In this case, the gas falls directly onto the disc edge. The associated velocity variation suggests that material is shocking in this region.

In Fig.~\ref{fig:beta_streamer}, we present a large-scale equatorial slice at $t=60 \, \mathrm{kyr}$ showing the plasma parameter $\beta_P$. Interestingly, the streamer and the central protostar-disc system are characterized by $\beta_P>1$, indicating that the gas in these regions is thermally-dominated. On the contrary, the surrounding low-density "bubble" is characterized by $\beta_P<1$, indicating that the gas is magnetically-dominated. This configuration points towards a magnetic origin for the streamer formation, such as the interchange instability proposed by \citet{mignon2021collapse} in the context of massive star formation or the magnetic process discussed in \citet{tu2023protostellar} for the collapse of a turbulent, low-mass protostellar core.

\subsection{Connecting to the central system : looking for shock signatures}
\label{subsec:shock_signature}
   \begin{figure}
   \centering
   \includegraphics[scale=1, width=\linewidth]{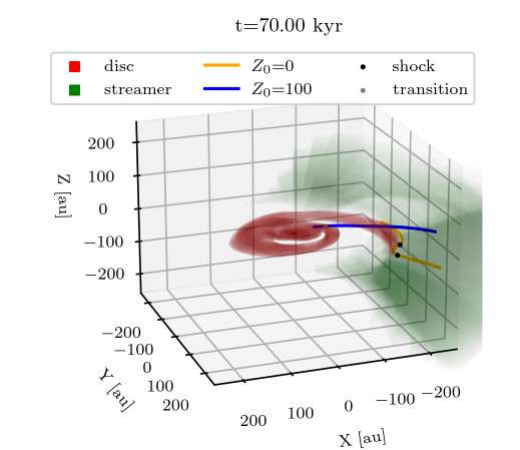}
   \caption{Three-dimensional representation of the disc and streamer with two attached streamlines at time $70 \, \mathrm{kyr}$. Red surfaces correspond to the disc, while green surfaces represent the streamer. Both streamlines start at $R=350 \, \mathrm{au}$ and $\varphi=\varphi_{streamer}$ with respectively $Z=0$ (orange solid line) and $Z=100 \, \mathrm{au}$ (blue solid line). Black dots indicate shocks (or rarefactions) along the streamlines while the grey dot corresponds to a density transition (see Fig.~\ref{fig:streamlinesShock} for their identification). For an animated version of the figure with a large-scale visualisation of the streamer, see \url{https://cloud.univ-grenoble-alpes.fr/s/ZNwHrbWg8A24Trb}.}
   \label{fig:streamlines3D}%
    \end{figure}
%

   \begin{figure}
   \centering
   \includegraphics[scale=1, width=\linewidth]{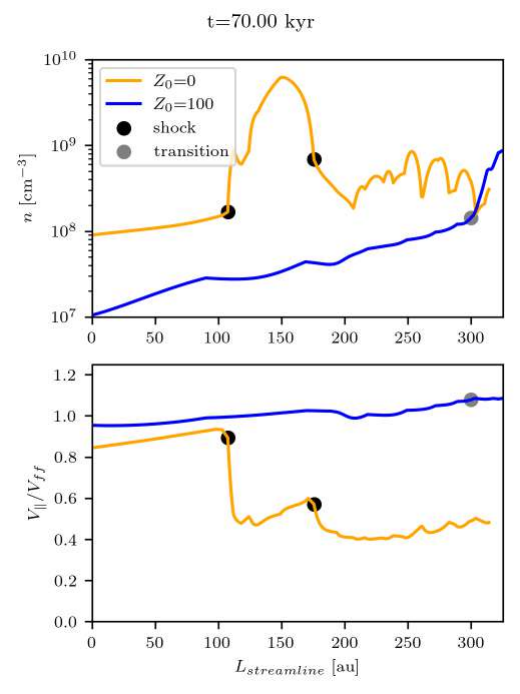}
   \caption{Top : particle density along the streamlines starting at $Z_0=0 \, \mathrm{au}$ and $Z_0=100 \, \mathrm{au}$ respectively. Bottom: gas velocity projected along each streamline. Colors and dots are the same as in Fig.~\ref{fig:streamlines3D}.}
   \label{fig:streamlinesShock}%
    \end{figure}

Figure \ref{fig:streamlines3D} provides a three-dimensional representation of the disc and streamer with two attached streamlines at time $70 \, \mathrm{kyr}$. The disc and streamer are displayed for representation purposes. In this plot, cells associated with the disc must have gas particle density over $10^9 \, \mathrm{cm^{-3}}$. The streamer corresponds to cells where $n \geq 10^6 \, \mathrm{cm^{-3}}$ and $u_r<0$ to ensure we focus on infalling material belonging to the parent molecular cloud. We additionally require $r>200 \, \mathrm{au}$ to exclude the central system. Streamlines are integrated with starting points of cylindrical radius $R_0=350 \, \mathrm{au}$ and azimuth $\varphi_0=\varphi_{streamer}$,
with respectively $Z_0=0$ and $Z_0=100 \, \mathrm{au}$ to probe the gas in the midplane and at elevated location.

The midplane streamline remains in the streamer and the midplane until it reaches the large-scale spiral arm where it is abruptly deflected (see Fig.~\ref{fig:streamlinesShock}) and entrained in the spiral motion. In contrast, the elevated streamline is channeled directly onto the innermost part of the disc, if not directly falling onto the seed (see the animated version of the plot for a better understanding).

The question of the shock is addressed by Fig.~\ref{fig:streamlinesShock}. The top panel shows the gas particle density in the streamline as a function of the position on the streamline while the bottom panel is the normalized gas velocity parallel to the streamline $V_\parallel/V_{ff}$.

For the midplane streamline, a first shock signature is observed at $100 \, \mathrm{au}$ with both discontinuities in density and velocity consistent with the encounter between the streamer and the spiral arm. The density jumps from roughly $10^{8}$ to almost $10^{10} \, \mathrm{cm^{-3}}$ and the gas loses more than half of its velocity. A fainter secondary rarefaction signature is observed around $170 \, \mathrm{au}$ where density and velocity drop again. We caution that the gas is not in a steady state, hence the streamline may not be representative of the gas kinematics after the first deflection. For the elevated streamline, we observe a sharp increase in density corresponding to the moment where the gas connects with the disc. However, the velocity along the streamline remains near free-fall at any position, and only a faint, smooth velocity gradient is observed. This corresponds to a smooth density transition, rather than a shock signature.

\subsection{Streamer mass and infall rate : impact on protostar-disc accretion}

To close our analysis of the streamer, we compute its mass and infall rate. We stay as close as possible to the computation method provided by \citet{valdivia2022prodige}.

We stick to the definition in Sect.~\ref{subsec:shock_signature} to flag cells belonging to the streamer. The mass is then computed by summing $\rho dV$ over each cell. We obtain $M_{streamer}\approx 0.02 \, \mathrm{M_\odot}$ corresponding to $3\mathrm{\%}$ of the protostar's mass and $10\mathrm{\%}$ of the disc mass at $70 \, \mathrm{kyr}$.

The infall rate $\dot{M}_{in}$ is the mass rate at which the streamer is infalling onto the protostar-disc system. It is not to be confused with the accretion rate $\dot{M}_\star$ directly onto the protostar. \citet{valdivia2022prodige} compute it through $\dot{M}_{in}=M_{streamer}/t_{ff,streamer}$ where $t_{ff,streamer}$ is computed using a streamline model. On each point of the streamline, they interpolate the velocity $V_\parallel$ and length variation $dl$. They can therefore compute the time needed to reach the disc.

We do the same using a streamline starting at $(R,Z,\varphi)=(R_{streamer},0,\varphi_{streamer})$, with $R_{streamer}\approx 1400 \, \mathrm{au}$ the outermost radius of the streamer. We then sum $dl/V_\parallel$ along the streamline to get $t_{ff,streamer}\approx 13 \, \mathrm{kyr}$ giving $\dot{M}_{in}\approx 1.5\times 10^{-6} \, \Macc$. Compared to the accretion rate $\dot{M}_\star\equiv \dot{M}(R=5 \, au) \approx 5\times 10^{-7} \, M_\odot \, yr^{-1}$ at the same time, we get a ratio of infall to accretion of $\dot{M}_{in}/\dot{M}_\star\approx 3$.

\section{Discussion}
\label{sec:discussion}

In this section, we confront our disc secular evolution and final GI-regulated state with observations. We also discuss the compatibility of our accretion streamer with what is observed by comparing its properties with the literature.

\subsection{Disc secular evolution and GI self-regulation}

In the first $15 \, \mathrm{kyr}$ of its life, the newborn disc is small, compact, and magnetized. It lies in the non-ideal regime and accretes through GI-driven spiral density waves. This is because efficient magnetic braking promotes accretion in the well-coupled MHD pseudo-disc, which in return increases the disc mass and makes it unstable. During the second phase, between $15$ and $45 \, \mathrm{kyr}$, the accretion in the pseudo-disc becomes less efficient and the disc can stabilize. In the meantime, the envelope provides high angular momentum material to the disc that can therefore expand. As a result, the accumulated magnetic flux is advected outwards and magnetisation decreases. In the third phase, lasting for the remaining $55 \, \mathrm{kyr}$, expansion is halted in the disc and mass accumulates at the edge from the pseudo-disc. At some point, the outer disc is massive enough to be unstable and the disc's final state is GI-regulated.

As a complement, we would like to emphasize an interesting result regarding the plasma parameter $\beta_P$. In subsection~\ref{subsec:secular_scenario}, we find that during phase II the disc is expanding, and its magnetic flux is advected along with the gas. In such a case, assuming the disc mass $M_d$ and poloidal magnetic flux $\phi_d$ to be constant, one can show that $\beta_P (R_d) \propto R_d^2$. This is verified in Fig.~\ref{fig:midplane_radial_prof}, bottom right panel. At $5\, \mathrm{kyr}$, $\beta_P(R_d)\approx 10$. Later on, once the disc radius has increased by roughly one order of magnitude, the associated plasma parameter is $\beta_P(R_d)\approx 10^3$. The verified dependency is a confirmation that once a given amount of flux is stored in the disc, it follows the gas evolution, and magnetisation evolves accordingly.

Numerically speaking, our collapse and disc's early evolution are in line with other works. Magnetic decoupling occurs by the time of the first core formation leading to a plateau with $B_z\simeq 100 \, \mathrm{mG}$ threading the disc, as expected in core-collapse simulations including ambipolar diffusion \citep{masson2016ambipolar,vaytet2018protostellar,xu2021formation2}. The initial disc size of roughly $10 \, \mathrm{au}$ is consistent with a magnetic regulation \citep{hennebelle2016magnetically}. A decreased magnetisation is then observed for simulations that properly resolve the disc vertical extent \citep{xu2021formation2}. The disc becomes massive and resistive enough to be gravitationally-regulated \citep{tomida2017grand,xu2021formation1,xu2021formation2}. The piling up of magnetic field at the transition between the diffusive and ideal MHD regimes is reminiscent of the magnetic wall proposed by \citet{li1996hydromagnetic} and observed also in \citet{xu2021formation2}, though with a fainter accumulation that could be explained by the differences in the diffusivity tables.

On the observational side, resolution is often missing to infer the surface density profile in class 0/I systems and we lack robust tracers to unveil the magnetisation. The few studies available for the surface density find a power-law index between $-1.7$ and $-1.5$ \citep{yen2014alma,aso2015alma,aso2017alma}. This is shallower than what we constrain to justify our GI-regulation mechanism ($\approx -2$). On the other hand, \citet{fiorellino2023mass} find that many class I have a disc-to-star mass ratio above 0.1, which they claim is typical for a GI-regulated disc. That being said, spirals are observed in only a few class 0/I protostars \citep{tobin2018vla,lee2020spiral}, while most of these objects do not show structures \citep{ohashi2023}. Yet, it is worth mentioning that these discs can be optically thick, in which case the spirals can be hidden.

The Class II stage is slightly better constrained. The review from \citet{andrews2020observations} summarizes class II disc properties inferred from observations. The constrained surface density profile has an index of $\approx -1$, again shallower than our finding, making GI regulation unlikely. Conversely, Zeeman measurements of the poloidal magnetic field give an upper limit of $\sim 1 \, \mathrm{mG}$ at a few $10 \, \mathrm{au}$ \citep{vlemmings2019stringent} consistent with our final disc magnetisation. 

Thus, most of the properties of our evolved disc are consistent with current observations, with the exception of the GI-regulated state, characterized by a steep surface density profile and prominent spiral density waves, which seem to be uncommon, even in young discs. The inclusion of internal heating, which is missing in the current model, could help stabilize the disc. Yet, it wouldn't change the final picture. Indeed, in our simulation, the triggering of GI is inevitable when magnetic braking is neutralized by an inefficient magnetic coupling because no other mechanism is able to evacuate the disc angular momentum as mass piles up.

This however ignores the role of MHD disc winds that could be launched from the ionised surface layers of the disc. In our simulation, a large-scale outflowing structure arises, but it is launched far from the disc surface. The importance of such outflows with elevated launching points is discussed in many core-collapse simulations \citep{machida2013evolution, tsukamoto2015effects, masson2016ambipolar,marchand2020protostellar} and they are proposed as a means to redistribute angular momentum. However, in our simulation, we find that the outflow is launched from a low density region (a "vacuum") which density is set by the numerical limiters (Alfv\'en and density floor). Therefore, in our view, its origin remains numerical.

Hence, we conclude that no proper MHD disc wind is found in our simulation, while many "disc only" simulations including surface ionisation exhibit MHD disc winds \citep{bethune2017global,bai2017hall,suriano2019formation,lesur2021systematic}. These surface ionisation processes are due to stellar far UV and X-ray photons that increase the ionisation fraction by orders of magnitude. These effects are absent in our simulation, in which we only consider cosmic ray ionisation in our chemical network. Hence it is tempting to associate the GI-regulated disc we obtain to the lack of stellar irradiation at the disc surface. This possibility should be investigated in the future.

\subsection{Large-scale streamer driving accretion in the envelope}

The long-term interaction of the envelope with the protostar-disc system leads to the formation of an accretion streamer. It is composed of near free-fall gas organized in a sheet-like configuration. It connects to the protostar-disc system either by shocking onto the disc edge in the midplane or by smoothly accreting onto the disc surfaces from higher altitudes. Quantitatively, the streamer maximum radius is $\approx 1400 \, \mathrm{au}$ with mass and infall rates of $0.02 \, \mathrm{M_\odot}$ and $1.5\times 10^{-6} \, \Macc$, corresponding to $3\mathrm{\%}$ of the seed mass and $3$ times the protostar accretion respectively.

Asymmetric large-scale structures are ubiquitous in numerical core-collapse models with sufficiently massive molecular clouds, accounting for turbulence \citep{kuffmeier2017zoom,lam2019disc, kuznetsova2019origins,wurster2019there,lebreuilly2021protoplanetary} or not \citep{mignon2021collapse}. The resulting envelope is often messy, exhibiting streamers with filamentary or sheet-like morphologies. \citet{tu2023protostellar} propose a magnetic origin of the streamer formation, which is also supported by \citet{mignon2021collapse} and consistent with our work.

Accretion streamers have been observed both in deeply-embedded class 0 \citep{pineda2020protostellar, murillo2022cold} and in class I sources \citep{yen2019hl, valdivia2022prodige, hsieh2023prodige}. They are rotating, free-falling, and connect to the disc \citep{pineda2020protostellar, valdivia2022prodige}. The meeting point is either associated with a smooth transition between the infalling streamer velocity and the Keplerian disc velocity \citep{yen2019hl,pineda2020protostellar} or it can present a sharp velocity drop consistent with shock tracing signatures \citep{ valdivia2022prodige}. It is remarkable that both these conclusions are in agreement with the kinematics in our streamer.

A large variety of streamer sizes have been observed  ($10^3-10^4 \, \mathrm{au}$). The streamer mass ranges between $10^{-3}$ and $10^{-1} \, \mathrm{M_\odot}$, and can be a significant fraction of the protostellar mass \citep[$0.1-10\mathrm{\%}$ of $M_\star$,][]{pineda2020protostellar,valdivia2022prodige,hsieh2023prodige}. The infall rate of the streamer is $\dot{M}_{in}=10^{-6} \, \Macc$, and can be of the same order as the protostellar accretion rate $\dot{M}_\star$, if not higher. \citet{pineda2020protostellar,valdivia2022prodige,hsieh2023prodige} find $\dot{M}_{in}/\dot{M}_\star\approx 1$, $5-10$ and $\geq 0.05$ respectively. Our computations are all lying in the observed range. 

\section{Conclusion}
\label{sec:conclusion}

We performed a 3D large timescale, non-ideal core collapse simulation in order to probe the secular evolution of embedded protoplanetary discs while paying specific attention to the magnetic field evolution and the disc's long-term interaction with the surrounding infalling envelope.

We follow the cloud collapse until the first hydrostatic core formation leading to disc settling and integrate for an additional $100 \, \mathrm{kyr}$. The simulation lasts for about $20 \, \mathrm{\%}$ of the class I stage \citep{evans2009spitzer}. Yet in the meantime, $90 \, \mathrm{\%}$ of the envelope is accreted by the seed or onto the disc, pointing towards an ending class I. This faster evolution of the numerical model with respect to the observations is probably a consequence of the dynamically unstable initial condition.

We achieve a resolution of $10$ cells per scale height (assuming $\epsilon=0.1$) in order to properly capture magnetic field diffusion, field line twisting, and GI-induced spirals. Our main results are:

\begin{enumerate}
    \item The disc experiences three accretion phases. In particular, once the disc has settled, magnetic braking mainly controls accretion in the pseudo-disc. Conversely, self-gravity controls angular momentum transport in the disc through spiral density waves triggered via Toomre instability. When gas is expanding, it is thanks to the envelope providing high angular momentum material through the disc surfaces.
    \item During phase II, the disc expands while keeping its mass and flux roughly constant. As a result, the plasma parameter at the disc edge follows $\beta_P(R_d)\propto R_d^2$. This dependency is evidence that the initial amount of magnetic flux is conserved throughout the disc evolution, and magnetisation evolves accordingly to the gas motion.
    \item The disc ends up in a GI-regulated state maintaining $R_\mathrm{d}$ around $60 \, \mathrm{au}$. Its surface density profile is shaped by a critical surface density radial profile of index $\approx -2$.
    \item The early evolution of the disc reproduces well the results from core-collapse models such as disc compacity, magnetic field decoupling, magnetic and later GI regulation. However, no MHD disc wind is found in our simulation. This is a natural outcome of efficient decoupling, yet it contrasts with disc-only models that systematically find MHD disc winds. We conjecture that this could be due to a lack of stellar ionisation processes.
    \item The expanding Keplerian gas and the decrease of the magnetic field qualitatively match class II observations. Yet, the observed power-law surface density is too steep to trigger gravitational instabilities, and the presence of Toomre-driven spirals is not supported by observations. 
    \item After $30 \, \mathrm{kyr}$, the envelope is organized in a large-scale, sheet-like accretion streamer that feeds the disc. It smoothly connects to its surfaces from elevated locations and shocks onto the outer rim in the midplane. It is a significant reservoir of mass whose infall rate is comparable to the accretion rate of the protostar.
\end{enumerate}

In conclusion, this secular run shows that the neutralisation of magnetic braking due to an efficient decoupling leads the disc to a nearly pure hydrodynamical state where GI is the only means to accrete. Consequently, the disc is stuck in a GI-regulated state shaping its surface density profile and final radius.

Yet, this scenario is difficult to support, due to the lack of observed spirals in embedded systems. This suggests that current models overestimate the importance of diffusion after disc formation. It would be interesting to explore additional ionisation processes susceptible to recovering a magnetically dominated accretion. The choice of initial conditions and the impact of grain size may also act upon the diffusions. This will be the topic of a forthcoming paper.

\begin{acknowledgements}
    We wish to thank our anonymous referee for valuable comments and suggestions that improved the present paper. We are thankful to Matthew Kunz, Benoit Commerçon and Patrick Hennebelle for fruitful discussions about the physics of collapsing cores. This project has received funding from the European Research Council (ERC) under the European Union’s Horizon 2020 research and innovation program (Grant Agreement No. 815559 (MHDiscs)). This work was supported by the "Programme National de Physique Stellaire" (PNPS), "Programme National Soleil-Terre" (PNST), "Programme National de Hautes Energies" (PNHE), and "Programme National de Planétologie" (PNP) of CNRS/INSU co-funded by CEA and CNES. This work was granted access to the HPC resources of IDRIS under the allocation 2023-A0140402231. Some of the computations presented in this paper were performed using the GRICAD infrastructure (https://gricad.univ-grenoble-alpes.fr), which is supported by Grenoble research communities.
    Data analysis and visualisation in the paper were conducted using the scientific Python ecosystem, including numpy \citep{Harris.ea20}, scipy \citep{Virtanen.ea20} and matplotlib \citep{Hunter.07}.
\end{acknowledgements}

%
%

\bibliographystyle{aa} 
\bibliography{ref.bib} 

\begin{appendix}    

\section{Self-gravity solver}
\label{app:sgsolver}

\subsection{Implementation in \textsc{Idefix} }

The \textsc{Idefix} code is upgraded with a self-gravity module. By resolving Eq.~(\ref{eq:poisson}), it infers the self-gravitational potential from the gas density distribution.

The Laplacian operator is discretized using second-order finite difference with self-consistent boundary conditions. The resulting matricial system is solved iteratively by a biconjugate gradient stabilized (BICGSTAB) algorithm. It uses the \textsc{Kokkos} routines encapsulated in \textsc{Idefix} to handle parallelisation \citep{Trott.Lebrun-Grandie.ea22,lesur2023idefix}.

A Jacobian preconditioner $P$ can be used to fasten convergence. It is designed as the diagonal part of the discretized Laplacian. It proved to be efficient at easing convergence when the grid is irregular.

While the BICGSTAB algorithm is the one used in the present work, we implemented two additional methods: a conjugate gradient (CG) and a minimum residual (MINRES). There is a loss of generality when switching from one to another: CG requires the operator to be symmetric positive definite, while MINRES only assumes symmetry and BICGSTAB has no constraint. On the other hand, improving generality increases computational cost and/or slows down convergence. Hence, the implementation of several solvers provides an optimum solution depending on the problem.

\subsection{"origin" boundary condition}
\label{app:origin}

To circumvent the problem of the singularity at the center of the grid, we implement a specific "origin" inner boundary condition for the self-gravity solver. It expands the grid radially inwards with a constant spacing so as to entirely fill the inner "hole" of the grid. We assume that the gas density is zero in this extension of the domain.

The Poisson equation for the gravitational potential is then solved by the self-gravity solver on this extended domain. Because the domain includes the origin, there is no need to prescribe any inner boundary condition in this approach.

\subsection{Validation tests}

   \begin{figure}
   \centering
   \includegraphics[scale=1]{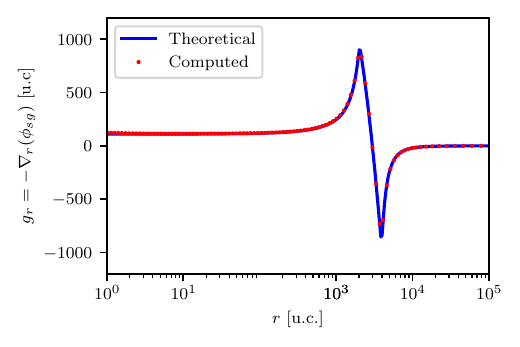}
   \caption{Static validation test: radial profile of the gravitational field along $(\theta,\varphi)=(\pi/2,0)$. The radius is in log scale. The grid configuration and boundary conditions are the same as our fiducial run, but we halved the resolution on each axis, uniformely for each patch. The density distribution is a uniform off-centered sphere of radius $1000$, located at $(r,\theta,\varphi)=(3000, \pi /2, 0)$. We set $\rho_0=\frac{3}{4\pi}$ and $G\equiv 1$ and the quantities are displayed in code units. The blue line is the theoretical profile, red dots are the computed data.}
    \label{fig:static}%
    \end{figure}
%

   \begin{figure}
   \centering
   \includegraphics[scale=1.]{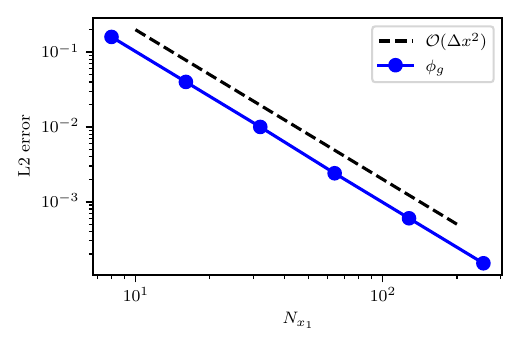}
   \caption{Convergence rate of the gravitational potential (blue dotted line) as a function of the grid spatial resolution using the BICGSTAB method. It is based on the off-centered sphere test. It exhibits second order spatial convergence.}
    \label{fig:order_accuracy}%
    \end{figure}
%

   \begin{figure}
   \centering
   \includegraphics[scale=1. ]{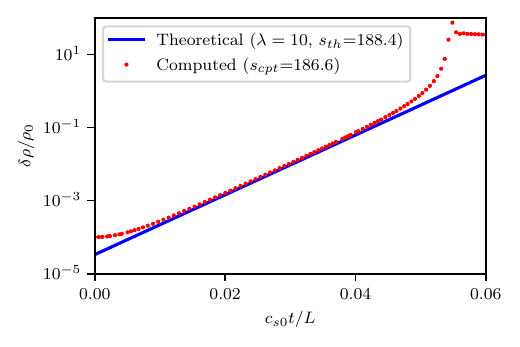}
   \caption{Dynamic validation test: amplitude of density fluctuations in log scale as a function of time following Jean's instability for $\lambda_J=1/3$ where Poisson equation is solved at every timestep. All quantities are dimensionless. The blue line is the theoretical prediction for the most unstable mode ($\lambda=10 \, \mathrm{u.c.}$), the red line is the computed result.}
              \label{fig:dynamic}%
    \end{figure}

We validate our implementation of self-gravity with two tests. The first one is a static test and confirms that the gravitational potential retrieved from the solver is accurate compared to the predicted one. The second is a dynamic test, where the dynamical solver is coupled with the self-gravity solver. It is based on Jean's instability and makes sure we properly capture mode growth.

Figure \ref{fig:static} shows the radial profile along $(\theta,\varphi)=(\pi/2,0)$ of the gravitational field in code units, inferred from an off-centered spherical, uniform density distribution. We compute the gravitational field rather than the potential to get rid of the integration constant and to make the comparison easier.

We took the same grid configuration and boundary conditions as our fiducial run. We halved the resolution on each axis, uniformely for each patch, in order to reduce the computation time while keeping the grid anisotropy. The density distribution is uniform inside an off-centered sphere of radius $1000\, \mathrm{au}$. The center is located at $(r,\theta,\varphi)=(3000, \pi /2, 0)$. We emphasize that only the self-gravity solver is tested here. Thus, as the physics is unimportant, we set $\rho_0=\frac{3}{4\pi}$ and $G\equiv 1$, and the quantities are displayed in code units. We set the convergence threshold to $10^{-5}$.

The theoretical and computed solutions are well matching thanks to the high resolution and low convergence threshold. The convergence rate for this test is about $600$ iterations, starting from a zero initial guess potential. After this first "burn-in" computation, the solver requires between 1 and $\sim 10^2$ iterations to converge, depending on the dynamics of the gas (it is $10$ on average for our fiducial run). We checked that the scheme is second-order accurate for the gravitational potential (see Fig.~\ref{fig:order_accuracy}).

Figure \ref{fig:dynamic} shows the amplitude of density fluctuations in log scale as a function of time following Jean's instability with the Poisson equation solved every timestep. Both quantities are adimensionned by background density $\rho_0$ and $c_{s0}/L$ respectively, $L=10 \, \mathrm{u.c.}$ being the domain size.

The setup is $1$D cartesian with periodic boundary conditions. The $x$ coordinate is meshed with $1000$ uniform cells and ranges between $0$ and $L$. The density distribution is initialized with a Gaussian perturbation of amplitude $10^{-4}$. Setup is adiabatic with $\gamma=5/3$, background density and pressure are $\rho_0=3$, $P_0= 1/\gamma$ in code units which  gives $\lambda_J=1/3$ ($G\equiv \pi$).

For a given wavelength $\lambda$, the expected growth rate is given by $s=2\pi (c_s/\lambda)\sqrt{|1-(\lambda/\lambda_J)^2|}$ with $\lambda>\lambda_J$. The mode of the largest wavelength is therefore the most unstable. Then, for $\lambda=L$, the theoretical growth rate is $s_{th}=188.4 \, c_s/L$, and the associated perturbation should dominate the evolution of density perturbation.

This is confirmed by red dots, associated with the computed evolution of density perturbation, which matches the theoretical linear prediction for the most unstable mode (blue solid line) where $c_{s0}t/L $ is in the range $0.1-0.4$. A linear regression in this portion of the slope gives $s_{cpt}=186.6$ corresponding to a relative error of $0.9\mathrm{\%}$. Hence, the dynamic is properly captured by our self-gravity solver.

\section{Gravity step}
\label{app:gstep}

The gravity calculation is performed just before the dynamical step. It triggers the gravitational potential computation from various sources. In our case, that includes self-gravity (see Appendix~\ref{app:sgsolver}) and point mass contribution.

In order to properly account for the \emph{whole} gravitational feedback, the missing inner seed is assimilated to a point mass with:

\begin{equation}
    \label{eq:pmpot}
    \phi_{pm} = -\frac{GM_{pm}}{r}
\end{equation}

\noindent
where $M_{pm}$ and $\phi_{pm}$ are respectively the mass and associated potential of the point mass.

The initial mass is the one enclosed in a uniform sphere of radius $r_{in}$ and density $\rho_0$. We sum up mass fluxes over the inner shell during the integration to update the central mass according to mass transits. The net gravitational potential used for the dynamical integration is then $\phi_g=\phi_{pm}+\phi_{sg}$.

One can specify the frequency of the gravity step. \citet{bethune2019envelopes} showed that updating the gravitational potential every $4$ dynamical timestep does not substantially impact the system evolution (see their test on Jeans' instability). 

We conducted our own test and obtained a relative error of $8\mathrm{\%}$ on the growth rate of the most unstable mode when computing gravity every $4$ dynamical timestep. We consider this variation acceptable and choose to compute gravity every $4$ timestep in our simulation to speed up the integration.

\section{Chemical network\label{app:chemNet}}

The magnetic diffusivities depend on the abundances of the charge
carriers. To compute these abundances, we consider a simple chemical
network based on \citet{umebayashi1990} and
\citet{kunz2009nonisothermal}. The network includes the following
reactions:

\begin{align}
  \mathrm{H_2}                      \xrightarrow{\mathrm{CR}} \mathrm{H_2^+}   + \mathrm{e^-}                                \label{eq:h2-ionisation} \\
  \mathrm{H_2^+} + \mathrm{H_2}     \rightarrow                       \mathrm{H_3^+}   + \mathrm{H}                          \label{eq:ion-neutral-1} \\                   
  \mathrm{H_3^+} + \mathrm{CO}      \rightarrow                       \mathrm{H_2}     + \mathrm{HCO^+}                      \label{eq:ion-neutral-2} \\
  \mathrm{Fe}    + \mathrm{HCO^+}   \rightarrow                       \mathrm{Fe^+}    + \mathrm{H}       + \mathrm{CO}      \label{eq:ion-neutral-3} \\
  \mathrm{HCO^+} + \mathrm{e^-}     \rightarrow                       \mathrm{H}       + \mathrm{CO}                         \label{eq:diss-recomb-1} \\
  \mathrm{Fe^+}  + \mathrm{e^-}     \rightarrow                       \mathrm{Fe}      + \mathrm{photon}                     \label{eq:rad-recomb-1} \\
  \mathrm{e^-}   + \mathrm{grain}   \rightarrow                       \mathrm{grain^-}                                       \label{eq:e-neutral-grain} \\                      
  \mathrm{e^-}   + \mathrm{grain^+} \rightarrow                       \mathrm{grain}                                         \label{eq:e-charged-grain} \\
  \mathrm{Fe^+}  + \mathrm{grain}   \rightarrow                       \mathrm{Fe}      + \mathrm{grain^+}                    \label{eq:ion-neutral-grain-1} \\
  \mathrm{Fe^+}  + \mathrm{grain^-} \rightarrow                       \mathrm{Fe}      + \mathrm{grain}                      \label{eq:ion-charged-grain-1} \\               
  \mathrm{HCO^+} + \mathrm{grain}   \rightarrow                       \mathrm{H}       + \mathrm{CO}      + \mathrm{grain^+} \label{eq:ion-neutral-grain-2} \\
  \mathrm{HCO^+} + \mathrm{grain^-} \rightarrow                       \mathrm{H}       + \mathrm{CO}      + \mathrm{grain}   \label{eq:ion-charged-grain-2} \\
  \mathrm{H}     + \mathrm{H}       \xrightarrow{\mathrm{grain}}      \mathrm{H_2}                                           \label{eq:h2-formation}
\end{align}

The ionization of $\mathrm{H_2}$ (Eq.~\ref{eq:h2-ionisation}) is
solely due to cosmic rays. We neglect shielding and focussing effects of cosmic rays, so the ionization rate $\zeta =
\mathrm{1.3 \times 10^{-17} \, \mathrm{s^{-1}}}$ is assumed to be constant. The reaction rates
for the ion-neutral
(Eqs.~\ref{eq:ion-neutral-1},~\ref{eq:ion-neutral-2}~and~\ref{eq:ion-neutral-3}),
the dissociative recombination (Eq.~\ref{eq:diss-recomb-1}), and the
radiative recombination (Eq.~\ref{eq:rad-recomb-1}) reactions are
given by:

\begin{equation}
  k = \alpha \, {\left( \frac{T}{300 \, \mathrm{K}} \right)}^\beta
\end{equation}

\noindent
where $T$ is the temperature and $\alpha$ and $\beta$ are the prefactor
and temperature exponent, respectively. The values of $\alpha$ and
$\beta$ for each reaction are given in Table~\ref{tab:reaction-network}.

\begin{table}
  \caption{Rate coefficients for the ion-neutral, dissociative recombination and radiative
    recombination reactions.\label{tab:reaction-network}}
  \begin{tabular}{lll}
    \hline
    \hline
    Reaction &$\alpha$ & $\beta$\\
             & $(\mathrm{cm^{3} \, s^{-_1}})$ & \\
    \hline
    Eq.~(\ref{eq:ion-neutral-1}) & $\mathrm{2.1\times10^{-9}}$  & 0     \\ 
    Eq.~(\ref{eq:ion-neutral-2}) & $\mathrm{1.7\times10^{-9}}$  & 0     \\ 
    Eq.~(\ref{eq:ion-neutral-3}) & $\mathrm{2.5\times10^{-9}}$  & 0     \\ 
    Eq.~(\ref{eq:diss-recomb-1}) & $\mathrm{2.0\times10^{-7}}$  & -0.75 \\ 
    Eq.~(\ref{eq:rad-recomb-1})  & $\mathrm{2.8\times10^{-12}}$ & -0.86 \\ 
    \hline
  \end{tabular}
\end{table}

The reaction rates for electron attachment
(Eq.~\ref{eq:e-neutral-grain}) and charge exchange reactions on
neutral grains
(Eqs.~\ref{eq:ion-neutral-grain-1}~and~\ref{eq:ion-neutral-grain-2})
are given by:

\begin{equation}
  k = \pi a^2 {\left( \frac{8 k_B T}{\pi m} \right)}^{1/2} \left[ 1 +
    {\left( \frac{\pi e^2}{2 a k_B T} \right)}^{1/2} \right] S
\end{equation}    

\noindent 
where $a$ is the grain radius, $ k_B$ is the Boltzmann constant, $m$ is
the mass of the electron or the ion, $e$ is the electron charge, and
$S$ is a sticking coefficient, assumed to be 0.6 for electrons and 1
for ions, respectively. For electron attachement
(Eq.~\ref{eq:e-charged-grain}) and charge exchange reactions on
charged grains
(Eqs.~\ref{eq:ion-charged-grain-1}~and~\ref{eq:ion-charged-grain-2}),
the reaction rates become:

\begin{equation}
  k = \pi a^2 {\left( \frac{8 k_B T}{\pi m} \right)}^{1/2} \left( 1 +
    \frac{e^2}{a k_B T} \right) \left[ 1 +
    {\left( \frac{2}{2 + \left( a k_B T / e^2 \right)} \right)}^{1/2}
  \right] S
\end{equation}

We assume that grains are spherical with a radius $a = 0.1
\,\mathrm{\mu m}$. The gas-to-dust mass ratio is assumed to be equal to
100. Assuming that the grains have a mass
density of $3 \, \mathrm{g \, cm^{-3}}$, this gives a grain abundance
  with respect to H nuclei of $\mathrm{1.3\times10^{-12}}$.

Finally, we assume the following reaction rate for the $\mathrm{H_2}$ on
grains (Eq.~\ref{eq:h2-formation}):

\begin{equation}
  k = \alpha {\left( \frac{T}{300 \, \mathrm{K}} \right)}^{1/2}
\end{equation}

\noindent
with $\alpha = \mathrm{4.95\times10^{-17}} \, \mathrm{s^{-1}}$.

Table~\ref{tab:initial-abundances} gives the initial abundances. We
assume that all hydrogen is in molecular form and that all carbon is
in the form of CO.\@ Iron is assumed to be ionized, so it has the
same initial abundance as free electrons. Grains are assumed to be
initially neutral. The abundances of all species in the chemical
network are computed as a function of time using \textsc{Astrochem}
\citep{maret2015astrochem}, until the steady-state equilibrium is
reached.

\begin{table}
  \caption{Initial abundances with respect to H nuclei of the species
    considered in chemical network.\label{tab:initial-abundances}}
  \begin{tabular}{ll}
    \hline
    \hline
    Species         & Abundance                    \\
    \hline
    $\mathrm{H_2}$  & 0.5                          \\
    $\mathrm{CO}$   & $\mathrm{8.4\times10^{-5}}$  \\
    $\mathrm{Fe^+}$ & $\mathrm{2.05\times10^{-6}}$ \\
    $\mathrm{e^-}$  & $\mathrm{2.05\times10^{-6}}$ \\
    \hline
  \end{tabular}
\end{table}

\begin{figure}
  \centering
  \includegraphics[width=\columnwidth]{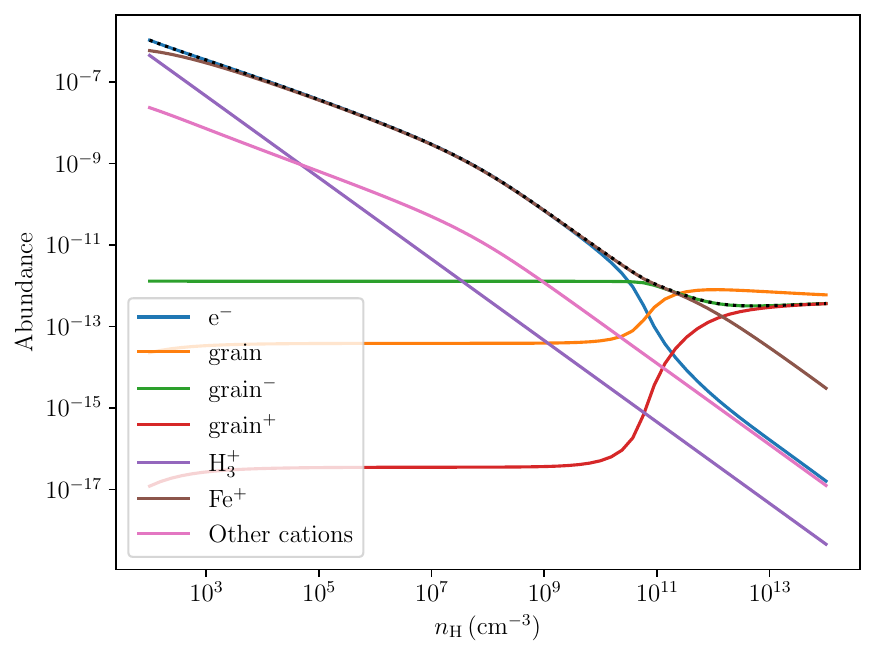}
  \caption{Abundances of the charge carriers at the steady-state as a
    function of the H number density for $a = 0.1 \,\mathrm{\mu m}$ and
    $\zeta = \mathrm{1.3 \times 10^{-17} \, s^{-1}}$. The black
 dotted line shows the ionization fraction.\label{fig:charge-carriers-vs-nh}}
\end{figure}

Fig.~\ref{fig:charge-carriers-vs-nh} shows the abundances of the main
charge carriers at the steady-state and the ionization fraction
(i.e. the total abundance of positively or negatively charged species
with respect to H nuclei), as function of the H number density, for a
grain radius of $a = 0.1 \,\mathrm{\mu m}$ and a gas temperature given
by Eq.~(\ref{eq:EOS_red}). The abundances of the main charge carriers
are in agreement with \citet[][see their Fig.~2, which corresponds to
the same grain size and the similar initial abundances that those
adopted here]{umebayashi1990}. The ionization fraction decreases with
the density for $ n_\mathrm{H} < 10^{11} \, \mathrm{cm^{-3}}$ and
remains constant at higher densities. For densities lower than
$10^{11} \, \mathrm{cm^{-3}}$, the main charge carriers are free
electrons and $\mathrm{Fe^+}$ ions. At higher densities, the main
charge carriers are positively and negatively charged grains. The
transition between the two regimes occurs when the ion fraction
becomes comparable to the number density of grains.

\begin{figure}
  \centering
  \includegraphics[width=\columnwidth]{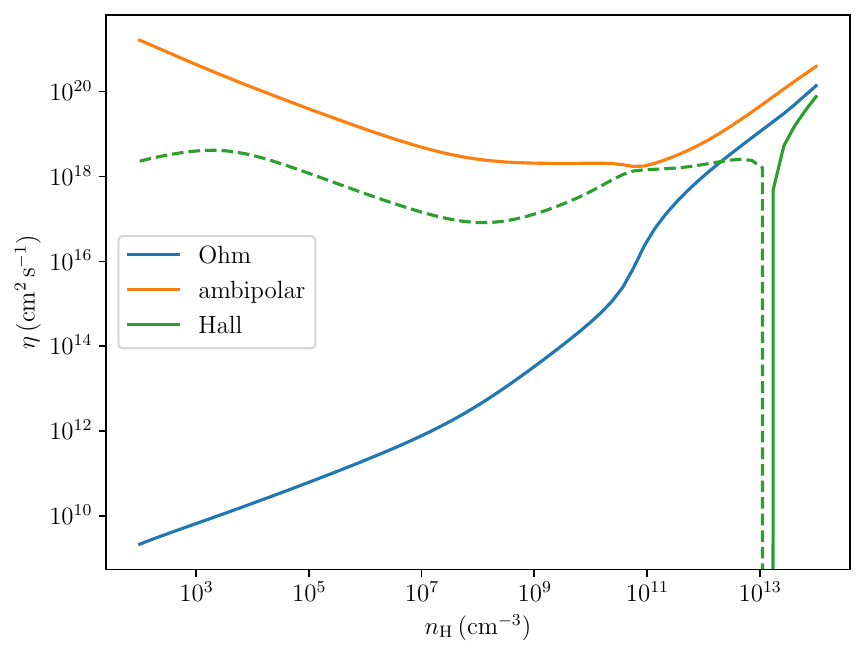}
  \caption{Magnetic diffusivities as a function of the H number
    density for $a = 0.1 \,\mathrm{\mu m}$, 
    $\zeta = \mathrm{1.3 \times 10^{-17} \, s^{-1}}$, and $B = 0.1
    \mathbf{\left( n_\mathrm{H} / \mathrm{cm^{-3}} \right)^{0.5}} \, \mathrm{\mu
      G}$. The dashed line corresponds to a negative
    diffusivities.\label{fig:diffusivities-vs-nh}}
\end{figure}

Fig.~\ref{fig:diffusivities-vs-nh} shows the corresponding magnetic
diffusivities \cite[see Eqs.~3.4, 3.5 and 3.6
in][]{lesur2021magnetohydrodynamics} as a function of the H number
density, for a magnetic field intensity $B = 0.1
\mathbf{\left( n_\mathrm{H} / \mathrm{cm^{-3}} \right) ^{0.5}} \, \mathrm{\mu G}$. The
diffusivities are in agreement with \citet[][see their
Fig.~C2]{xu2021formation1}.

\section{Internal boundaries}
\label{app:intBdy}

   \begin{figure}
   \centering
   \includegraphics[scale=1, width=\linewidth]{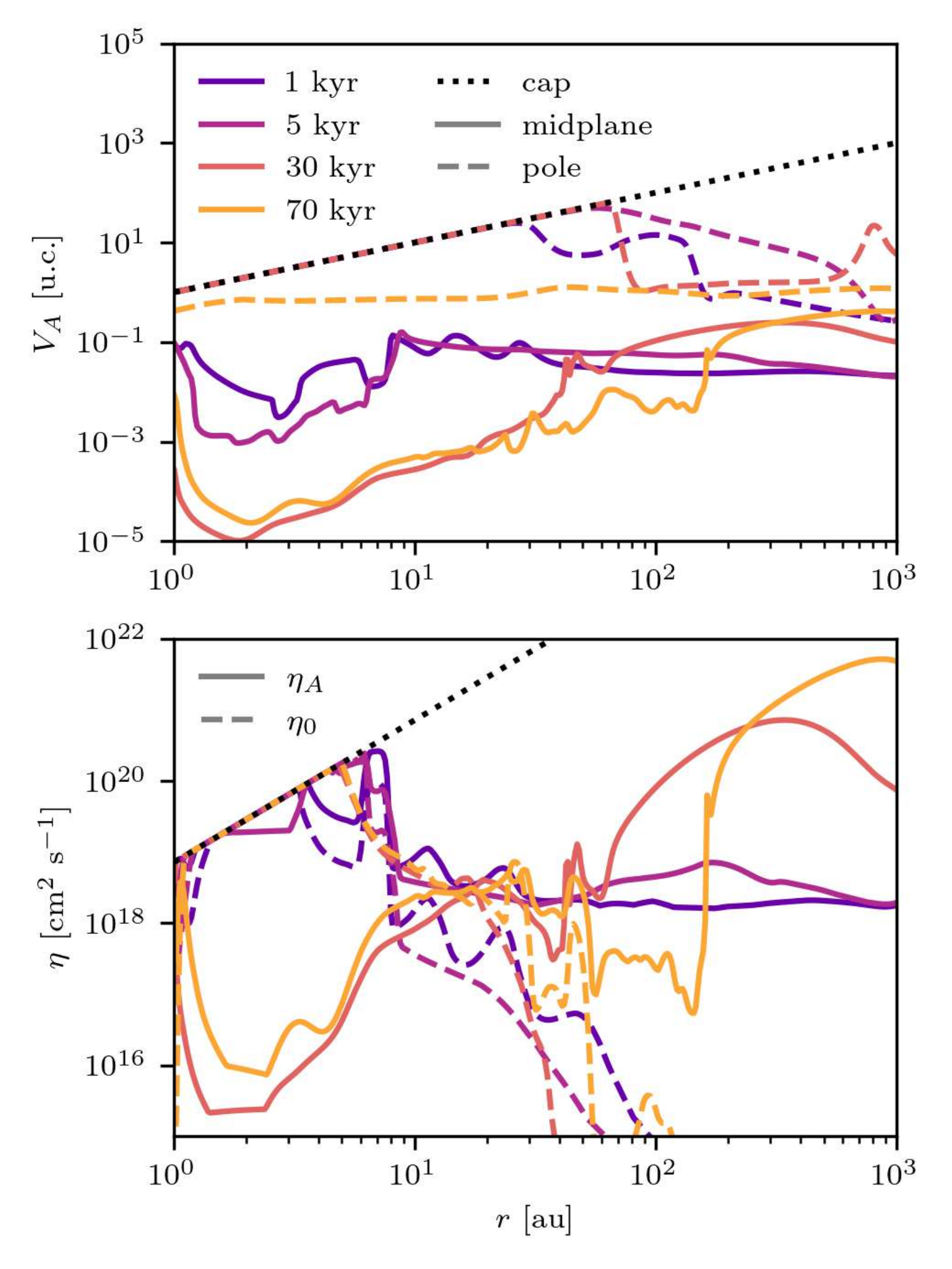}
   \caption{Top: Alvén speed in code units versus the radius, where $\varphi=0$. Solid lines are the midplane values ($\theta\approx \pi/2$) while dashed lines are closer to the pole ($\theta\approx \pi/3$). The lighter the color, the later the snapshot. The black dotted line is the Alvén cap $V_A=V_{A,max}r$ where $V_{A,max}=1 \, \mathrm{u.c.}$. Bottom: ambipolar (solid) and Ohmic (dashed) diffusions versus the radius. Color coding and black dotted line are the same as for the top panel, with the diffusivity cap $\eta=\eta_0 r^2$ where $\eta_0\approx 7.1\times 10^{18} \, \mathrm{cm^2 \, s^{-1}}$.}
              \label{fig:caps}%
    \end{figure}

We use three internal boundaries in order to prevent a dramatic drop of the timestep without significant loss of accuracy: an Alvén speed limiter, diffusivity caps, and an advection timestep limiter.

In code units, Alvén speed is defined  by

\begin{equation}
    \label{eq:alven}
    V_A = \frac{B_{\textsc{Idefix}}}{\sqrt{\rho}}
\end{equation}

 \noindent
where $B_{\textsc{Idefix}}\equiv B/\sqrt{4\pi}$. 

Consequently, in strongly magnetized, low-density regions it can become very high and require very low timesteps, incompatible with the large timescale integration. 

To alleviate this problem we introduce an Alvén speed limiter: for any cell of radius $r$ if $V_A>V_{A,max}r$ the density is replaced with $\rho_{new}=B_{\textsc{Idefix}}^2/(V_{A,max}r)^2$. The associated velocities are updated following $u_{i,new}=u_i\cdot \rho/\rho_{new}$ to satisfy as much as possible momentum conservation. Only $u_\varphi$ is left untouched. In this simulation, we set $V_{A,max}$ to $1$ in code units.

Figure \ref{fig:caps}, the top panel presents the Alvén speed profile at $\varphi=0$ in the midplane (solid lines) and near the pole (dashed lines) for snapshots of increasing time. In the midplane, except for the very beginning, the Alvén speed limiter is never triggered. This is not the case near the pole. Because of cavity carving, the areas above and below the seed are strongly magnetized with low density and we need to limit the Alvén speed up to $100 \, \mathrm{au}$. That being said, the cavity region is barely discussed in this work because it is poorly described in the actual framework.

We also use diffusivity caps following \citet{xu2021formation1,xu2021formation2}. The timestep associated with diffusion processes is proportional to $\Delta l^2/\eta$, where $\Delta l$ is the typical cell size and $\eta$ is the diffusivity coefficient associated with the diffusion process. Particularly strong values of $\eta_A$ and $\eta_0$ are therefore susceptible to dramatically slow down the integration.

We solve the problem by introducing a diffusivity cap such that for any cell of radius $r$, if $\eta_{A,O}>\eta_0 r^2$, the diffusivity coefficient is replaced with $\eta_{A,O}=\eta_0 r^2$. Here, $\eta_0 \, \approx 7.1\times 10^{18} \, \mathrm{cm^2 \, s^{-1}}$ is the conversion factor from code units to physical units. 

Figure \ref{fig:caps}, bottom panel shows the ambipolar (solid lines) and Ohmic (dashed lines) diffusivity profiles at $\varphi=0$ for the same snapshots as the top panel. The cap is triggered for ambipolar diffusion only in the very beginning. Ohmic diffusion, however, is limited for radii below $5 \, \mathrm{au}$ as soon as the disc forms and remains so until the end. This is expected as Ohmic diffusion increases with gas density.

The last internal boundary, the advection timestep limiter, is a consequence of the first one. In the innermost regions, gas affected by the Alvén limiter can't be launched in the outflow. Conversely, it is falling back onto the seed, reaching high velocities that limit the timestep. To solve this problem, we use a timestep limiter such that where $dt_{adv}<dt_{min}$, the velocity components are updated following $u_{i,new}=u_i\cdot dt_{min}/dt_{adv}$. We set $dt_{min}=1 \, \mathrm{u.c.}$

We monitored the total mass and angular momentum in the system during the integration to ensure that none of these routines was significantly affecting their balance.

\end{appendix}

\end{document}